\documentclass[manuscript,screen]{acmart}

\AtBeginDocument{
  }

\setcopyright{acmlicensed}
\copyrightyear{2026}
\acmYear{2026}
\acmDOI{XXXXXXX.XXXXXXX}
\acmConference[]{ACM Trans. Comput.-Hum. Interact.}{June 03--05, 2025}{Woodstock, NY}
\acmISBN{978-1-4503-XXXX-X/2018/06}

\usepackage[caption=false,font=normalsize,labelfont=sf,textfont=sf]{subfig}
\newcounter{RQCounter}
\newcommand{\AddRQ}[1]{\textbf{\refstepcounter{RQCounter}\label{#1}RQ\arabic{RQCounter}:}}

\usepackage{tabu}                      
\usepackage{booktabs}                 
\usepackage{multirow}
\newcommand{\Romannum}[1]{\uppercase\expandafter{\romannumeral#1\relax}}
\newcommand{\romannum}[1]{\expandafter{\romannumeral#1\relax}}
\usepackage{comment}
\DeclareUnicodeCharacter{2212}{-}
\usepackage{tabularx}   
\newcolumntype{Y}{>{\centering\arraybackslash}X}

\renewcommand\footnotetextcopyrightpermission[1]{}
\authorsaddresses{} 

\begin{document}

\fancyhead[LE,RO]{}
\fancyhead[RE,LO]{}

\fancypagestyle{firstpagestyle}{
  \fancyhead[LE,RO]{}
  \fancyhead[RE,LO]{}
  \fancyfoot[RO,LE]{}
  \fancyfoot[RE,LO]{}
  \fancyfoot[C]{}
}

\fancypagestyle{standardpagestyle}{
  \fancyfoot[RO,LE]{}
  \fancyfoot[RE,LO]{}
  \fancyfoot[C]{}
}
\pagestyle{standardpagestyle}

\title{ErgoGlide: A Wearable Trackball Device for Ergonomic Text Entry in Virtual Reality}

\author{Muhammad Abu Bakar}
\email{abubakar.nuaa@gmail.com}
\orcid{0000-0003-1160-8524}
\author{Yu-Ting Tsai}
\email{hieicis91@hotmail.com}
\orcid{0000-0002-4087-594X}
\author{Muhammad Imran}
\email{imranghawind@gmail.com}
\orcid{0009-0002-3382-9934}
\affiliation{%
  \institution{Yuan Ze University}
  \city{Taoyuan}
  \country{Taiwan}
}

\author{Yan-Ann Chen}
\email{chenya@ntut.edu.tw}
\orcid{0000-0002-3348-6022}
\affiliation{%
  \institution{National Taipei University of Technology}
  \city{Taipei}
  \country{Taiwan}
}

\begin{abstract}
In virtual reality, it is challenging to achieve satisfactory text entry speed/accuracy, ergonomics, usability, and learnability.
To address this issue, we developed ErgoGlide, a novel lightweight and compact wearable device that facilitates text entry tasks in virtual environments.
The proposed ErgoGlide can be regarded as a small trackball that is wearable on a user’s finger like a ring.
By using ErgoGlide with a hive-like virtual keyboard, the user can rotate the ball for key selections, making text entry intuitive and accurate.
We conducted three user studies to evaluate ErgoGlide and found that key confirmation techniques have significant effects on text entry speed and the hive-like keyboard design significantly reduced thumb movements.
Furthermore, ErgoGlide can significantly improve typing accuracy, ergonomics, and usability over previous text entry methods.
Experimental results also indicated that the typing speed of ErgoGlide can be notably improved after training.
\end{abstract}

\begin{teaserfigure}
    \centering
    \begin{tabu}{c@{\hspace{0.1cm}}c@{\hspace{0.1cm}}c}
      \multirow{2}{*}{\subfloat[Mixed-reality visualization of our system]{\includegraphics[height=5.2cm]{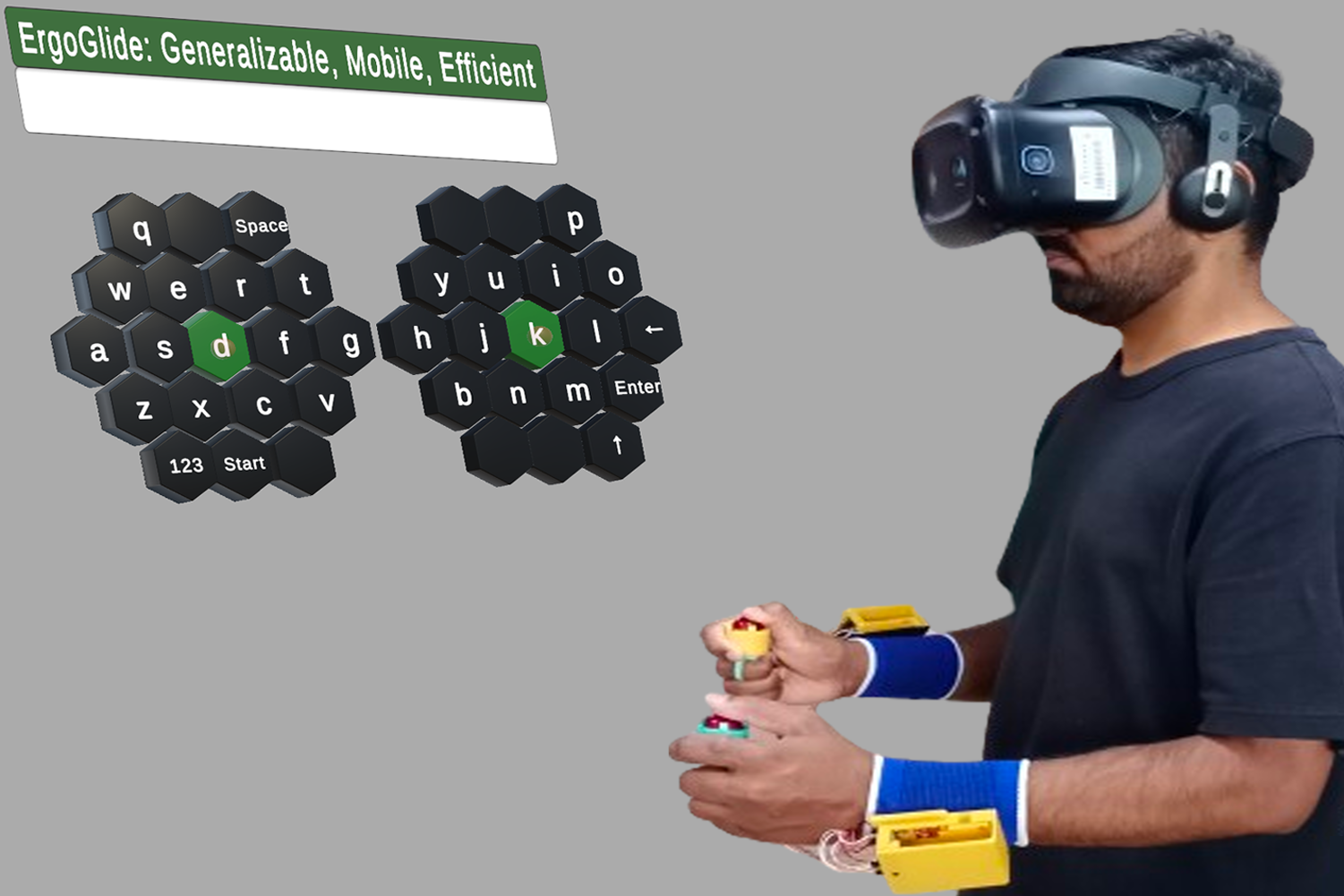}}} & 
      \multirow{2}{*}{\subfloat[Assembly views]{\includegraphics[height=5.2cm]{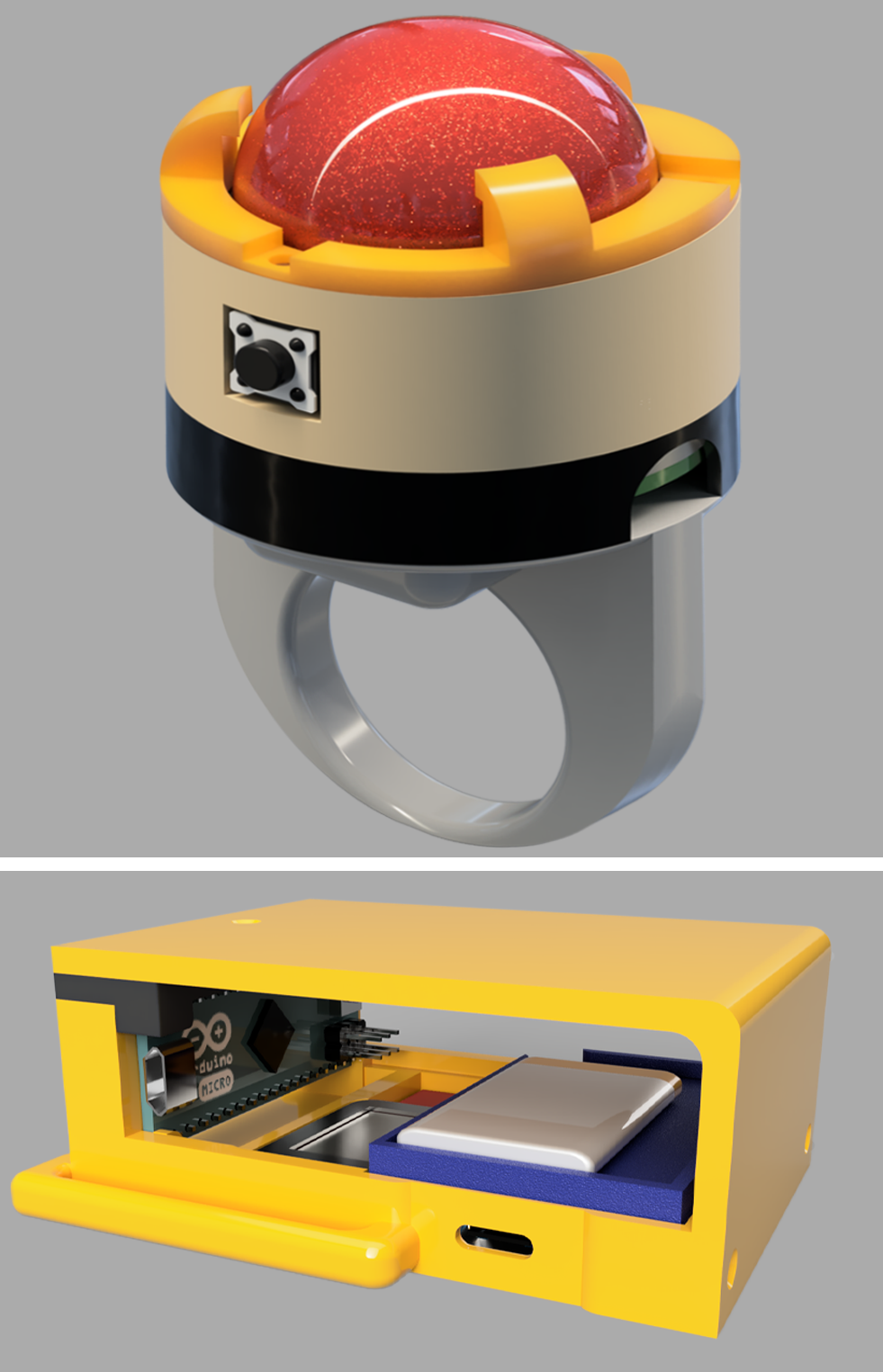}}} &
      \subfloat[Alphabetical mode]{\includegraphics[height=2.0cm]{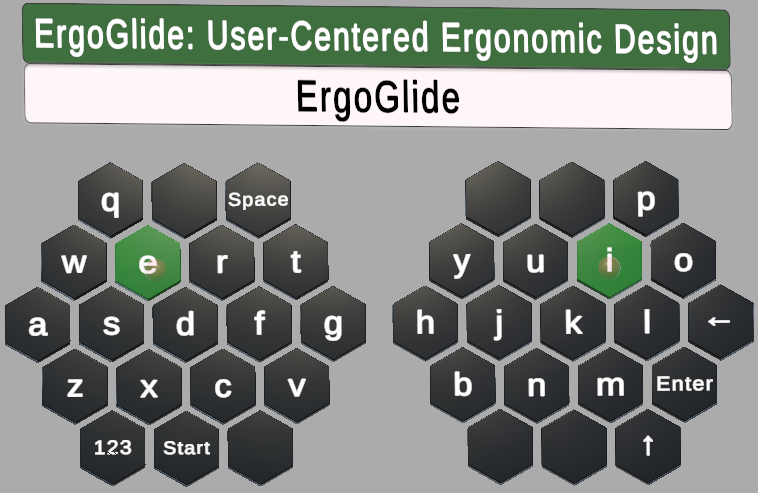}}\\
      && \subfloat[Symbolic mode]{\includegraphics[height=2.22cm]{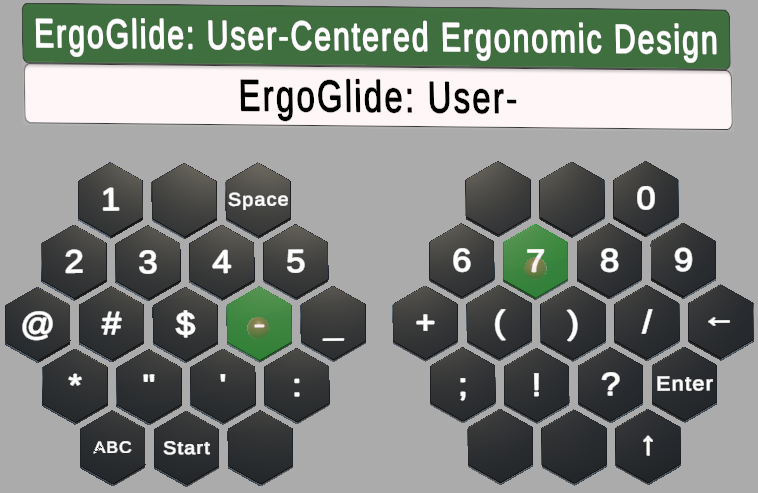}}
    \end{tabu}
    \caption{ErgoGlide is a lightweight and compact wearable device that supports ergonomic text entry in virtual \mbox{environments.} (a) A user performs text entry with ErgoGlide in a standing posture. The user can rotate the ball on ErgoGlide with their thumb to change the selected key, and press the button on ErgoGlide to enter the current key. (b) The assembly views of finger-worn (top) and wrist-mounted (bottom) parts of ErgoGlide. (c)-(d) Different modes of a hive-like virtual keyboard employed with ErgoGlide.}
    \Description{Teaser figure of ErgoGlide.}
    \label{Fig:teaser}
  \end{teaserfigure}

\maketitle
\section{Introduction}
Text entry devices have vastly evolved in the reality-virtuality continuum to facilitate various extended reality (XR) applications~\cite{ControllerTap}, such as social virtual reality (VR), web browsing, email writing, document editing, and so on.
Although a physical keyboard is efficient and popular for immobile and on-desk scenarios~\cite{PhysicalKeyboardinVR2,PhysicalKeyboardinVR}, it may not be technically suitable for mobile and confined spaces~\cite{ConfinedSpace}.
Other existing solutions are typically based on handheld~\cite{CrowbarLimbs,ControllerTap,FlowerText,PizzaText} and wearable devices~\cite{Velcro,RotoSwype,ThumbText}.
However, handheld devices are often designed for a specific XR system and do not support cross-system text entry~\cite{RotoSwype}.
On the other hand, wearable devices based on touch surfaces~\cite{ThumbText,FingerText,TipText} may suffer from imprecise selections due to the limited interaction area~\cite{Velcro,PrinType}.
Moreover, wearable devices based on gestures~\cite{RotoSwype,GestureGaze} may cause physical fatigue in user's hands and arms from frequent and large hand movements~\cite{TapGazer, TelemetRing}.
Another downside of these approaches is their incompatibility to type out-of-vocabulary words, such as special symbols and passwords~\cite{HandwritingusinController2}.
Therefore, developing a lightweight and compact device to support low-effort and cross-system text entry with mobility in immersive virtual environments remains an open challenge.

To address the above-mentioned problems, we present ErgoGlide (Figure~\ref{Fig:teaser}), a wearable trackball-based device that supports low-effort and ergonomic text entry while the user moves around.
The user can rotate the ball on ErgoGlide with their thumb to move a cursor continuously in the virtual environment for changing the selected key and press the button on the device to confirm key selections.
When compared to previous text entry devices, such as touch surfaces~\cite{FanPad} and thumbsticks~\cite{PizzaText}, rotating the ball on ErgoGlide can be low-effort and comfortable.
Furthermore, we developed a hive-like virtual keyboard that was specifically tailored for ErgoGlide.
This keyboard design aims to allow less overall thumb moving distance during prolonged use of ErgoGlide.

Moreover, we conducted three user studies (all approved by the research ethics committee) to evaluate the performance of ErgoGlide and explore more appropriate configurations.
The first user study was conducted to assess the impact of different key confirmation techniques and keyboard designs.
We found that pressing the button on ErgoGlide to confirm key selections is more intuitive than shaking and tapping, and a hive-like keyboard can significantly reduce thumb moving distances.
In the second user study, we compared ErgoGlide with FanPad~\cite{FanPad}, PizzaText~\cite{PizzaText}, and a variant of ErgoGlide with thumbsticks (namely "JoyGlide").
Experimental results indicated that ErgoGlide is superior to the other three in terms of ergonomics, usability, and text entry speed. 
To further investigate the learnability of ErgoGlide, we conducted a five-day user study with seven participants. Experimental results showed that during the last day, participants could achieve a mean text entry speed of 18.92 words per minute (WPM).

In summary, this paper makes the following contributions:

\begin{itemize}
\item We developed a trackball-based and ring-like device, namely ErgoGlide, to enable low-effort, ergonomic, and cross-system text entry with mobility in immersive virtual environments.
\item We designed a hive-like virtual keyboard to reduce the overall thumb moving distance.
\item We conducted three user studies to evaluate the effectiveness of ErgoGlide, compare it with three other methods, determine its appropriate configurations, including key confirmation techniques and keyboard designs, and investigate the performance impact of more training sessions on ErgoGlide.
\end{itemize}

ErgoGlide is specially designed for sitting, standing, and walking around scenarios where ergonomics hold great importance, such as the VR office~\cite{VROffice}.
VR programming is another practical scenario that requires accurate text entry for effective coding and debugging.
Without an efficient text entry method, VR developers may repetitively remove head-mounted displays (HMDs) to test and debug their programs throughout the day.
Additionally, ErgoGlide allows users to type text with small body movements, which are crucial for ergonomics~\cite{CrowbarLimbs}.
We believe that ErgoGlide has the potential to enable ergonomic, easy-to-learn, and efficient text entry for XR continuum.

\section{Related Work}
A large number of text entry methods for XR systems have been proposed and developed over time.
However, this section only presents a brief overview of previous approaches that are more related to ErgoGlide.
A comprehensive survey is beyond the scope of this paper.
Interested readers may refer to~\cite{VRTextEntryReviewPaper,Selection}.

\subsection{Wearable Devices}
Wearable text entry devices have received increasing attention due to their diverse applicability in the reality-virtuality continuum.
For example, some finger-worn devices, such as TypeAnywhere~\cite{TypeAnywhere}, Shapeshifter~\cite{Shapeshifter}, and QwertyRing~\cite{QwertyRing}, allow text entry by detecting finger taps/movements on a solid surface.
By contrast, ErgoGlide is a self-contained device that does not need to work with additional physical surfaces, thus increasing the mobility of users while using XR systems.

Moreover, RotoSwype~\cite{RotoSwype} facilitates gesture-based mid-air text entry in mobile scenarios by tracking hand movements using a finger-worn inertial measurement unit (IMU).
ThumbText~\cite{ThumbText} employs a finger-worn miniature touchpad to support a two-step text entry procedure.
A user slides their thumb on the touch surface and confirms the selection by releasing their thumb on the desired key.
Nevertheless, ThumbText may suffer from imprecise selections due to the limited interaction area of its miniature touchpad~\cite{Velcro,PrinType}.
In contrast to RotoSwype and ThumbText, ErgoGlide is a non-mid-air approach that does not rely on any gestures or touch surfaces with limited area.
Users also can intuitively and precisely rotate the ball to select a virtual key with low error rates.

Wrist-worn (also known as freehand) devices~\cite{TapID,TapType,GestureGaze} are another widely adopted category of wearable text entry hardware.
Streli et al.~\cite{TapType} developed a wristband by improving TapID~\cite{TapID} with more accurate IMUs and employed a Bayesian deep classifier to detect low-intensity finger taps, thus allowing efficient typing for trained expert users.
Moreover, Zhao et al.~\cite{GestureGaze} applied a wrist IMU device to speed up text entry by integrating gestures and gaze.
By contrast, ErgoGlide is a finger-worn and self-contained device that does not need an additional solid surface for haptic feedback.

There are also other wearable devices that map keys onto nails/fingers and leverage thumb-to-finger touch interaction to enable subtle text entry~\cite{FingerText,TipText}.
While these devices support on-the-go scenarios, their small interaction area often leads to a limited number of keys~\cite{PrinType}.
Additionally, some researchers have explored the usability and feasibility of hand-worn devices, such as gloves, for text entry~\cite{DigiTouch,Glove2,Glove4}.
However, hand interactions with the real world may be encumbered while wearing these devices.
By contrast, while wearing ErgoGlide, user's hands still can be employed for other interaction tasks.

\subsection{Non-Wearable Devices}
\textbf{Handheld controllers} offer versatile and flexible solutions to text entry in XR environments~\cite{CrowbarLimbs,HiPad,ControllerTap,ControllerHandWriting,FlowerText,Selection,AlphanumericTextEntry,FanPad}.
Kern et al.~\cite{ControllerTap} evaluated the performance of controller-based word-gesture and tapping techniques in augmented and virtual reality.
They discovered that tapping is superior to word-gesture in both environments.
Abu Bakar et al.~\cite{CrowbarLimbs} proposed a crowbar-like metaphor to reduce physical fatigue for long-term text entry tasks.
Spericher et al.~\cite{Selection} investigated six selection-based text entry techniques and found that pointing with controllers outperformed the others.
To facilitate efficient alphanumeric and special character entry, Wan et al.~\cite{AlphanumericTextEntry} presented two controller-based switching techniques for seamless transitions between different key modes.
Leng et al.~\cite{FlowerText} designed specialized keyboard layouts and utilized 3D translations of a controller to select characters.
Although controller-based text entry methods have become more and more popular in VR, they often do not support cross-system text entry.

\textbf{Bare-hand} techniques eliminate the need of handheld controllers and have recently showed a lot of potential for efficient text entry~\cite{BH1,VISAR}.
To support bimanual text entry with all fingers on a physical surface, TouchInsight~\cite{TouchInsight} incorporated a bivariate Gaussian distribution to address the egocentric hand tracking uncertainties of moving HMD cameras.
Dudley et al.~\cite{Dudley} exploited the hand tracking function of the HMD to investigate mid-air typing on a virtual QWERTY keyboard.
Shen et al.~\cite{3DGestureTrajectories} introduced a novel neural network based gesture-to-text decoding approach to translate three-dimensional trajectories of mid-air gestures into intended text.

\textbf{Handwriting:}
In recent years, researchers have investigated the potential of using controllers for direct handwriting in XR environments. 
Kern et al.~\cite{ControllerHandWriting} proposed to hold the controller in a pen-like posture and investigated handwriting on mid-air and physically aligned surfaces.
Fourrier et al.~\cite{HandwritingusinController2} explored the influence of virtual board orientations (slanted and vertical) with sensory feedback for mid-air text entry.
Their findings indicated that the slanted board combined with haptic, visual, and auditory cues offered advantages over other conditions.

\textbf{Input devices:}
Researchers have also focused on developing novel text entry hardware devices.
A cubic-shaped text entry device~\cite{KeyCube} with physical keys demonstrated that moving beyond traditional flat keyboard form factors can improve mobility and versatility in virtual and immersive workspaces.
Other than dedicated text entry devices, prior work has additionally employed traditional input devices that were not originally designed for text entry. For example, thumbsticks have been widely investigated for selection- and stroke-based approaches~\cite{JoyStick4,JoyStick2,PizzaText}.
These techniques require users to draw multiple strokes to enter a character, which may not be a seamless and efficient way.
Furthermore, Wobbrock and Myers proposed stroke-based text entry methods using a trackball~\cite{Trackball1} and presented a word completion extension~\cite{Trackball2}.
To the best of our knowledge, trackballs have never been applied to text entry tasks in XR environments.

\textbf{Gaze-based} techniques have been reported promising in combination with other approaches~\cite{GazaCombination3,GazaCombination1,GazaCombination4,GazaCombination5,GazaCombination2}.
Ren et al.~\cite{EyeHandTyping} integrated eye-tracking information to assist mid-air finger-based text entry, which outperformed the simple gaze-based technique.
Moreover, TapGazer~\cite{TapGazer} enables text entry by tapping on a surface and selecting desired words with gaze.
To eliminate dwell delays, Hu et al.~\cite{SkiMR} proposed a dwell-free technique with a statistical decoder to translate eye movements into text.

\textbf{Physical and virtual keyboards} have been studied in depth for text entry in virtual environments.
Researchers have investigated different aspects of physical keyboards, such as hand representations~\cite{HR1,HR2,HR3}, keyboard reconfiguration~\cite{PhysicalKeyboardinVR}, and performance analysis~\cite{PKPerformance}. 
Although physical keyboards allow fast text entry speed, mobility remains a challenging issue.
Furthermore, Dudley et al.~\cite{VRKeyboard} evaluated the performance of using index fingers or all ten fingers for typing with a virtual keyboard in mid-air or on a physically-aligned surface. They found that using only index fingers could lead to better performance in both cases.

\subsection{Circular Keyboards in XR}
When combined with a specialized input device, circular keyboards can provide additional advantages.
To facilitate single-handed text entry, Jian et al.~\cite{HiPad} mapped an alphabetically organized circular layout onto the touchpad of a VR controller.
Wu et al.~\cite{FanPad} instead employed two touchpads with various split QWERTY layouts and evaluated their performance.
However, these approaches are particularly designed for circular touchpads and may not be suitable for touchpads with different shapes~\cite{FanPad}.
To provide hands-free text entry, RingText~\cite{RingText} combines head movements with a circular keyboard, but may not be suitable for a user with restricted head movements.
Frequent head movements in VR environments are also prone to motion sickness.

\section{Design Rationale}
The proposed ErgoGlide was developed with specific objectives:
\begin{itemize}
\item \textbf{Comfort}: Facilitate low-effort and ergonomic text entry tasks.
\item \textbf{Efficiency and learnability}: Ensure adequate text entry speed and accuracy with a low learning curve.
\item \textbf{Mobility}: Support various postures/scenarios, such as sitting, standing, and walking around.
\item \textbf{Bimanual interaction}: Enable text entry with user's both hands.
\item \textbf{Directional flexibility}: Allow flexible and precise user control over key selections. 
\item \textbf{Cross-system interoperability}: Provide a general-purpose pointing device without dependencies on any specific VR systems/platforms.
\end{itemize}

\subsection{Comfort, Efficiency, and Learnability}
Some previous text entry methods/devices~\cite{HiPad,ThumbText,PizzaText} require users to frequently execute multiple steps for locating/entering a single character, which may not be intuitive and ergonomically user-friendly.
By contrast, ErgoGlide is specifically designed to facilitate low-effort and ergonomic text entry.
By rotating the ball to control a cursor for key selections, our device can remain stationary, which involves small thumb movements and does not require users to move other body parts.
Thus, physical fatigue can be reduced, especially for long-term text entry tasks.

Inspired by on-screen keyboards, ErgoGlide employs a similar two-step interaction process to first continuously move a cursor onto a target key and then confirm the selection by pressing the button on the device.
Furthermore, the keyboard design and layout may positively/negatively influence typing performance~\cite{LayoutImpact}.
The QWERTY keyboard layout has been a standard and user-familiar one over the past decades~\cite{QWERTY_Review}.
Researchers have also suggested it for VR text entry tasks~\cite{FlowerText,Selection}.
ErgoGlide thus adopts a keyboard layout based on QWERTY (Figure \ref{Fig:teaser}(c)) to achieve adequate typing efficiency and learnability with improved ergonomics.

\subsection{Bimanual Interaction and Mobility}
To enable bimanual and intuitive text entry~\cite{bimanual,PizzaText}, we follow previous work~\cite{FlowerText} to divide the virtual keyboard into two parts, each of which can be associated with one hand, and employ a layout similar to QWERTY.
Moreover, the lightweight and compact form factor of ErgoGlide allows users to wear and use it while walking around or in a sitting/standing posture.
This adaptability is an essential element in XR environments, where users are often required to move around.

\subsection{Directional Flexibility and Cross-System Interoperability}
Inspired by the intrinsic rolling characteristics of trackballs, we developed ErgoGlide to allow intuitive interaction with the virtual environment.
Compared with previous thumbstick-based approaches, the ball on ErgoGlide is omnidirectional, which is crucial for ergonomics, usability, and precise cursor control.
This directional flexibility especially allows users to freely rotate the ball in all directions to quickly move a cursor onto a specific location for accurate key selections.

Additionally, modern interactive XR applications often require universally applicable devices that everyone can use everywhere across different systems/platforms~\cite{CrossDevice}.
However, many prior text entry devices in XR do not support cross-system interoperability, for example, handheld VR controllers.
By contrast, ErgoGlide can work as a general-purpose pointing device without dependencies on specific systems/platforms.

\begin{figure}[t]
  \centering
  \subfloat[Finger-worn part]{\includegraphics[width=0.4\columnwidth]{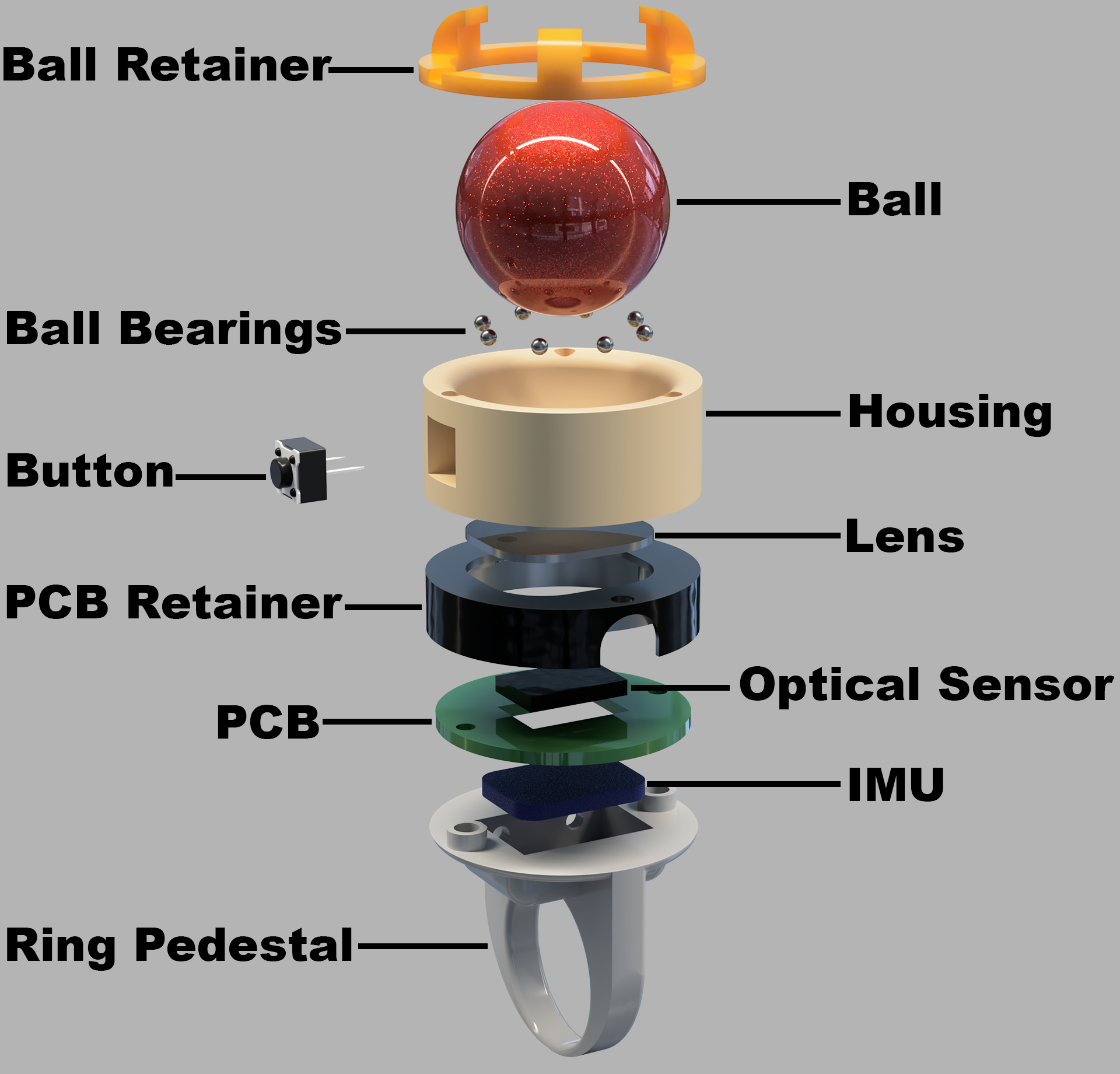}}\
   \hspace{0.3cm} 
  \subfloat[Wrist-mounted part]{\includegraphics[width=0.4\columnwidth]{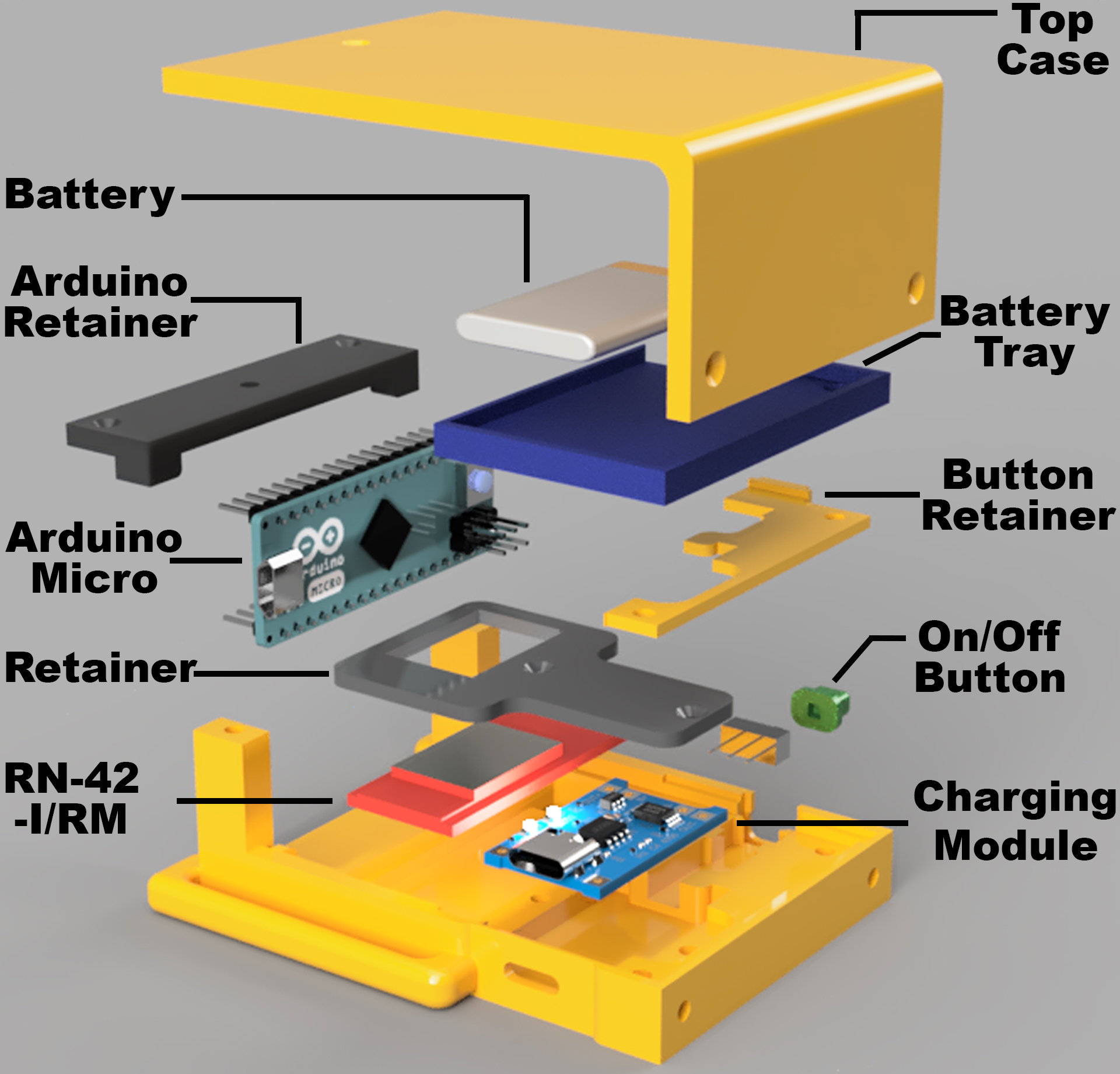}} 
  \caption{Exploded view diagrams of ErgoGlide components.}
  \Description{(a) Exploded view diagram of finger-worn part. (b) Exploded view diagram of wrist-mounted part.}
  \label{fig:FingerandWrist}
\end{figure}

\section{System Design}
Our system was developed based on the following hardware and software components:
\begin{itemize}
  \item ErgoGlide: A trackball-based and ring-like device for ergonomic user interactions.
  \item Hive keyboard: A hive-like virtual keyboard tailored for ErgoGlide.
\end{itemize}

\subsection{ErgoGlide}
As illustrated in Figure~\ref{fig:FingerandWrist}, ErgoGlide consists of the finger-worn and wrist-mounted parts.
The finger-worn part was mainly built from off-the-shelf components, including a ball, a housing with eight small steel bearings, an optical lens, an optical displacement sensor, an IMU, a tactile button, a printed circuit board (PCB), a PCB retainer, and a ring-like pedestal.
It weighs 73.6 grams (48.3 grams for the ball and 25.3 grams for all the others), with a height of 37 millimeters and an outer diameter of 33 millimeters.
The ball with a diameter of 25.4 millimeters was made of stainless steel and coated with phenolic resin.
The housing, the PCB retainer, and the ring pedestal were made of polylactic acid and created by a 3D printer (FlashForge Creator Pro 2).
The eight bearings (each with a diameter of 2 millimeters) in the housing were applied to carry the ball and facilitate its rotations.
We adopted a PMW3360DM-T2QU optical displacement sensor with a LM19-LSI lens from PixArt Imaging to measure two-dimensional displacement vectors that result from rotations of the ball.
Both the sensor and the lens were soldered on a PCB and connected to an Arduino Micro development board (with an ATMEGA32u4 microcontroller) through wires.
Moreover, a BMI160 IMU from Bosch Sensortec was employed to detect the tapping and shaking operations by using its built-in interrupt-based gesture detection. Additionally, a tactile button was also integrated for efficient key confirmation.

The Arduino board was placed in the wrist-mounted part and transmitted the measured displacement data to a personal computer via a RN42-I/RM Bluetooth module from Roving Networks, which operated at a baud rate of 115200 bits per second.
All electronic modules were powered by a YK-302040 lithium polymer battery (3.7 volt and 180 milliampere per hour) with a TP4056 charging module.
Note that other components of the wrist-mounted part were also created by the 3D printer.
Finally, the wrist-mounted part (58.7 grams) was attached to the wrist/forearm of a user by an elastic support.

\begin{figure}[t] 
  \centering
  \subfloat[Button]{\includegraphics[width=0.24\textwidth]{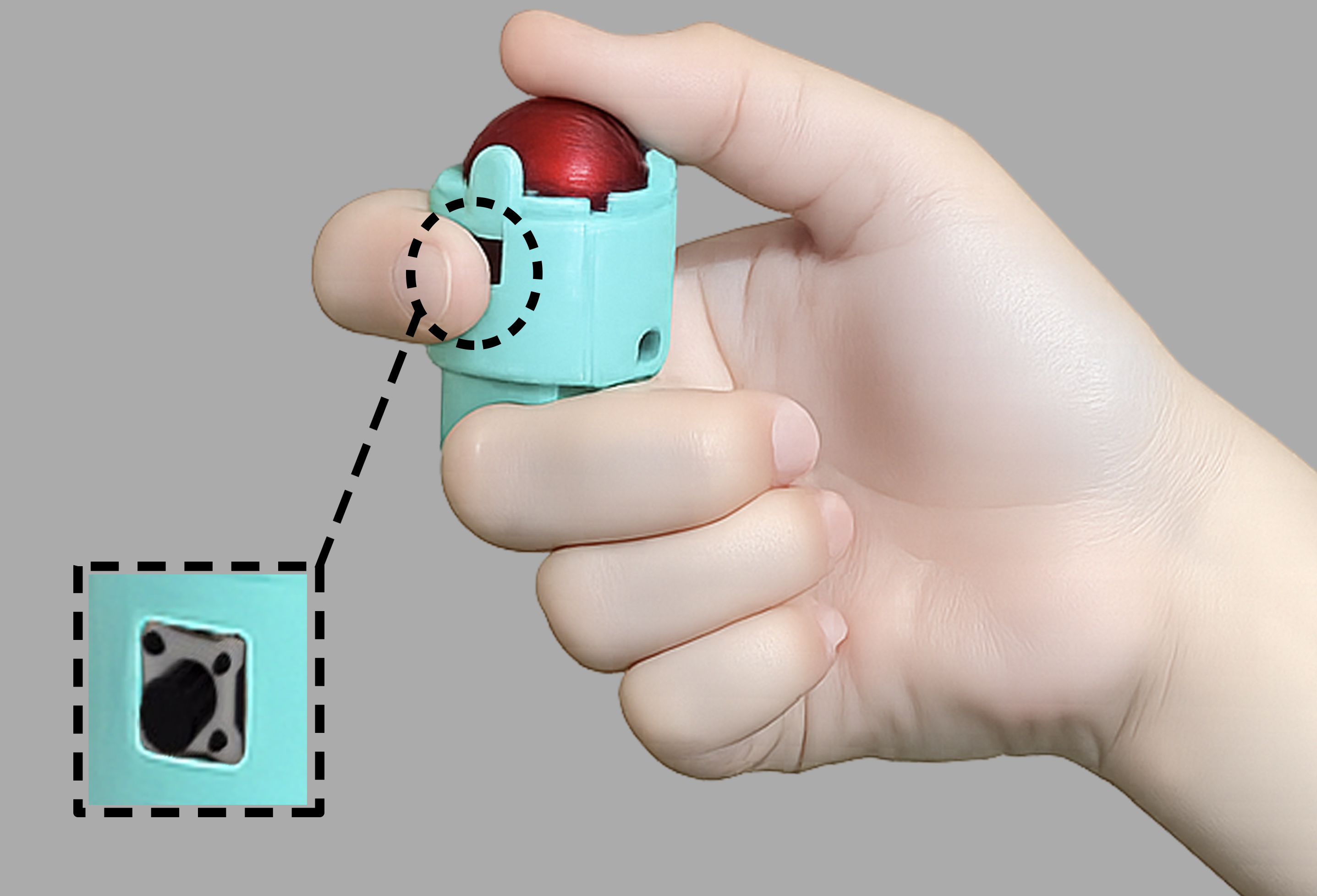}}\
  \subfloat[Shake]{\includegraphics[width=0.24\textwidth]{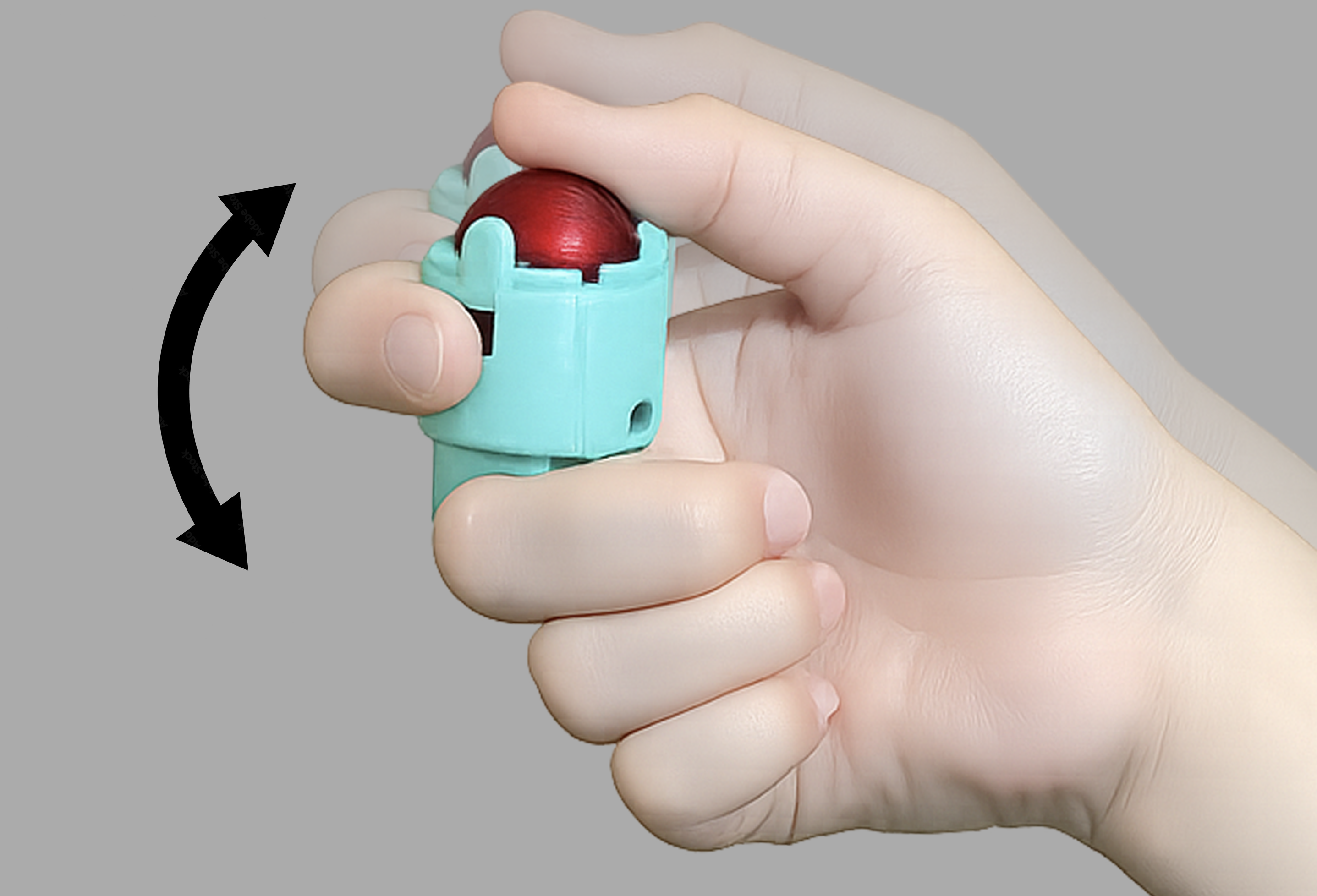}}\
  \subfloat[Tap]{\includegraphics[width=0.24\textwidth]{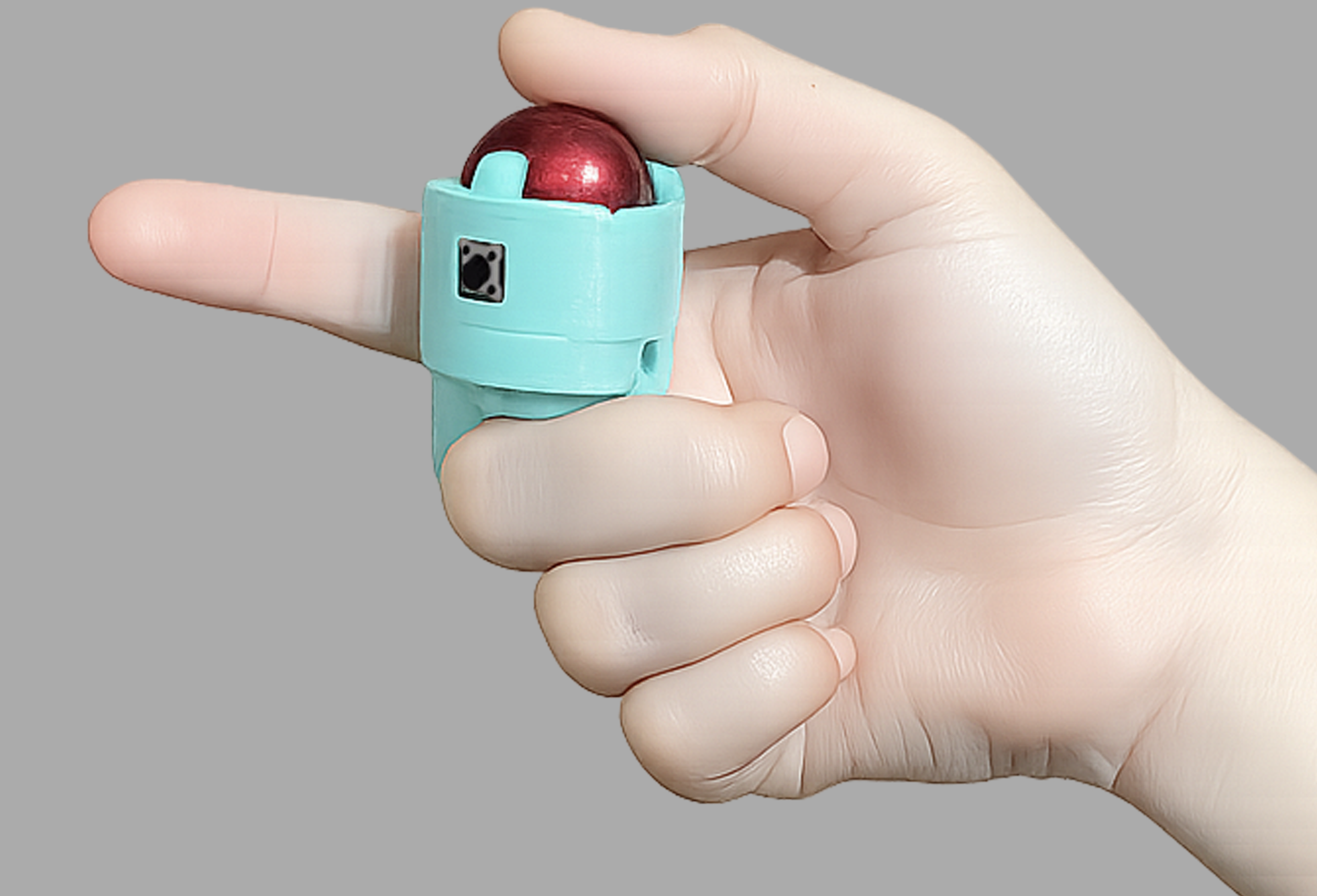}
  \includegraphics[width=0.24\textwidth]{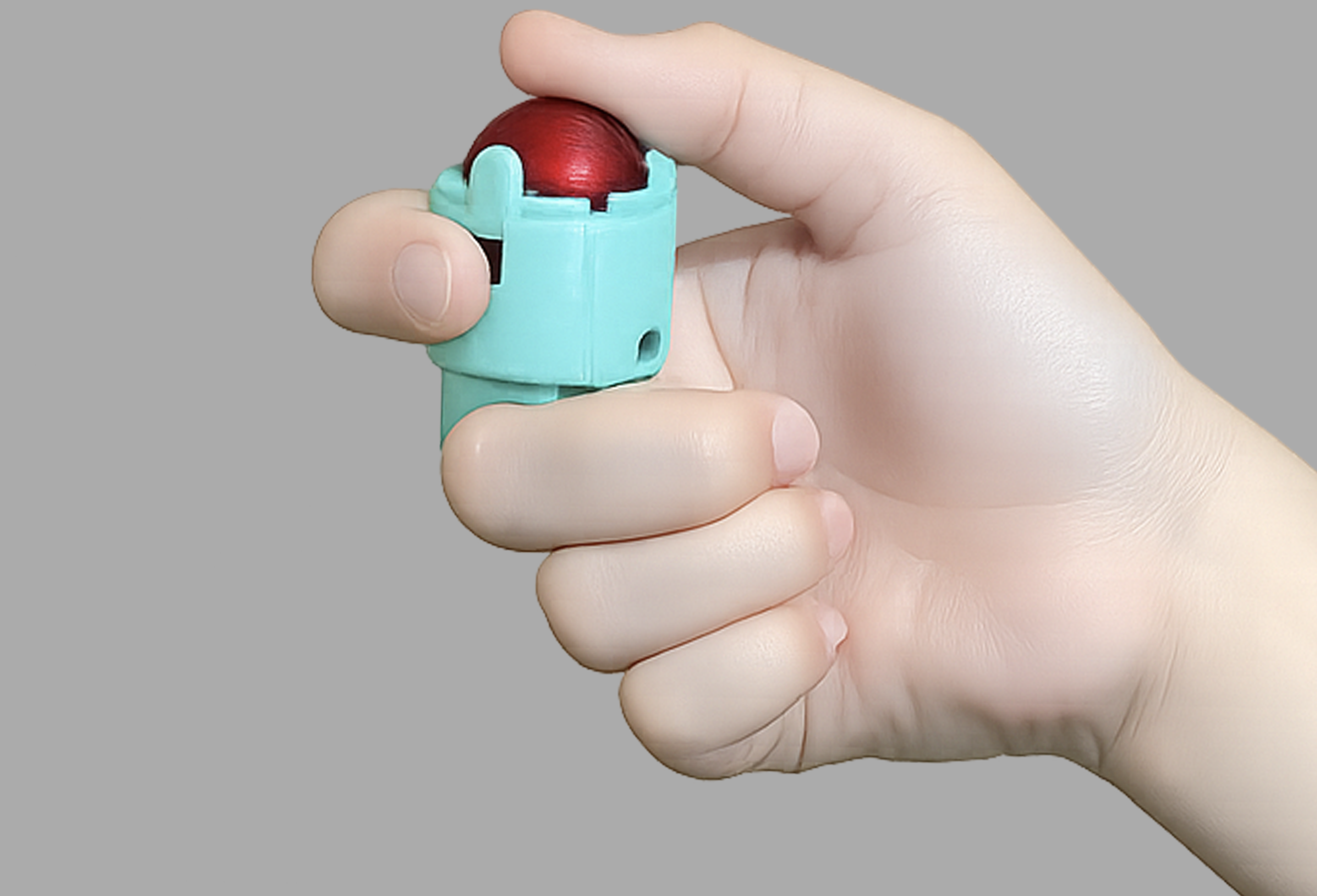}}
  \caption{ Different key confirmation techniques in User Study \Romannum{1}.
  (a) Pressing the tactile button with the index finger.
  (b) Shaking ErgoGlide down and up once.
  (c) Tapping ErgoGlide with the index finger.}
  \Description{Three key confirmation technique for ErgoGlide.}
  \label{fig:TapShake}
\end{figure}

\begin{figure}
  \centering
  \subfloat[Hive]{\includegraphics[width=0.4\columnwidth]{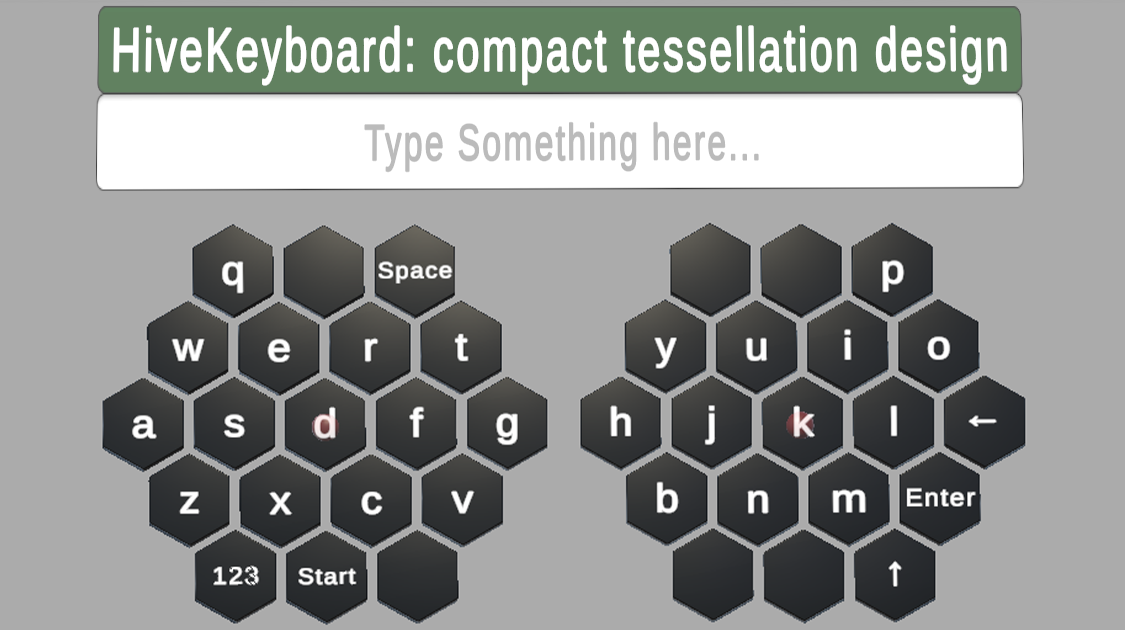}}\
  \hspace{0.3cm} 
  \subfloat[SplitQWERTY]{\includegraphics[trim=0in 1.5in 0in 0in, clip=true, width=0.52\columnwidth]{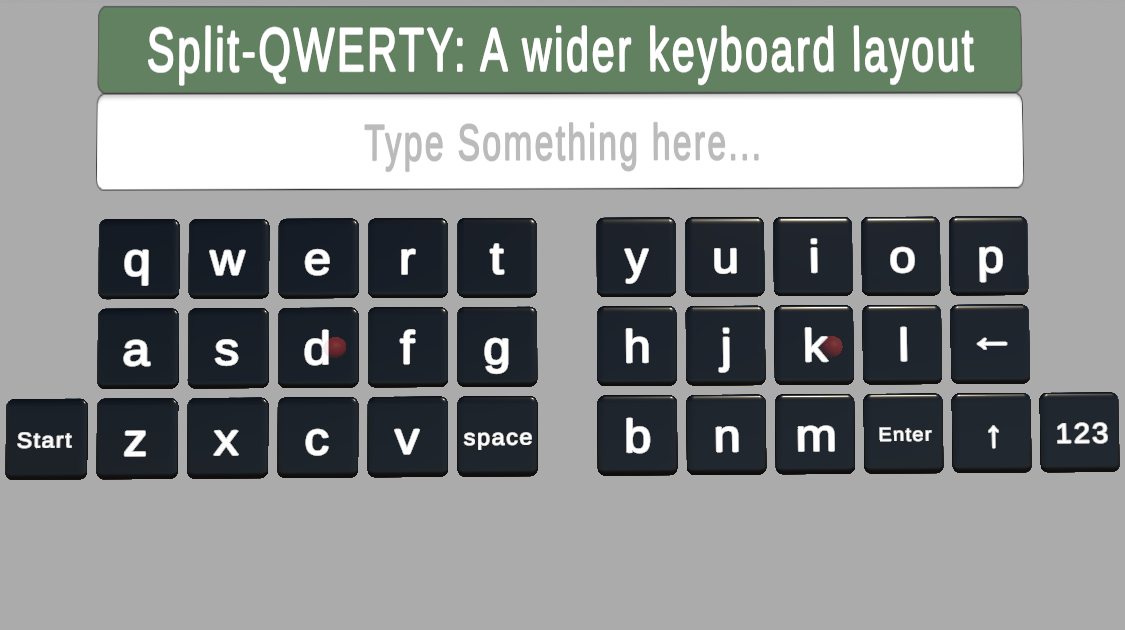}}
  \caption{Different keyboard designs in User Study \Romannum{1}.}
  \Description{Hive and SplitQWERTY keyboard employed with ErgoGlide.}
  \label{fig:KeyboardDesigns}
\end{figure}

\subsection{Hive Keyboard}
As demonstrated in Figure~\ref{Fig:teaser}(c) and Figure~\ref{Fig:teaser}(d), the virtual hive-like keyboard is divided into two parts, each of which is associated with an ErgoGlide device and one hand.
In each part, every key is represented by a hexagon and arranged in a compact hive pattern to facilitate a larger key size and more number of keys without widening the keyboard.
A user is supposed to wear an ErgoGlide device on each of their middle fingers to ensure workload balance between hands.
They can rotate the ball with their thumb to continuously move a cursor based on the measured two-dimensional displacement vectors and then change the selected key on the associated keyboard.
Note that we retain the common QWERTY layout, which often improves the learning curve, since most users are familiar with it.

\section{User Study \Romannum{1}: Key Confirmation Techniques and Keyboard Designs of ErgoGlide}
It was reported that typing performance and usability may be affected by key confirmation techniques and keyboard designs~\cite{KBDesign_ST_Impact}. 
To further improve the proposed ErgoGlide, we conducted this user study to compare three confirmation techniques (respectively called "Button", "Shake", and "Tap" in Figure~\ref{fig:TapShake}) and two keyboard designs ("Hive" and "SplitQWERTY" in Figure~\ref{fig:KeyboardDesigns}).
As the name suggests, the Button technique relies on pressing a tactile button to confirm a key selection (Figure~\ref{fig:TapShake}(a)).
Moreover, Shake and Tap techniques respectively confirm a key selection by shaking ErgoGlide down and up once (Figure~\ref{fig:TapShake}(b)) or tapping ErgoGlide with the index finger (Figure~\ref{fig:TapShake}(c)).
Note that both operations can be easily and accurately detected by the IMU on ErgoGlide.

\subsection{Research Questions}
Three key research questions for ErgoGlide were investigated in this user study:
\begin{itemize}
  \item \AddRQ{RQ1} Will key confirmation techniques and keyboard designs have significant effects on ergonomics, text entry speed/accuracy, and usability?
  \item \AddRQ{RQ2} Will Hive+Button, Hive+Shake, and Hive+Tap have similar speed/accuracy/usability/learnability and outperform other conditions?
  \item \AddRQ{RQ3} Will the Hive keyboard design reduce the overall thumb moving distances when compared to SplitQWERTY?
\end{itemize}

\subsection{Participants}
Eighteen participants (8 females and 10 males), between the ages of 20-43 ($M=25.56$, $SD=5.04$), voluntarily participated in this user study.
All were university students with diverse academic and ethnic backgrounds.
Nine participants had some experience with VR, five had limited experience with using trackballs, and nine had no experience with any of the aforementioned devices.
Moreover, two participants had previously engaged in VR-related text entry experiments.

\begin{figure}[t]
  \centering
  \subfloat[Mean Fatigue]{\includegraphics[width=0.45\textwidth]{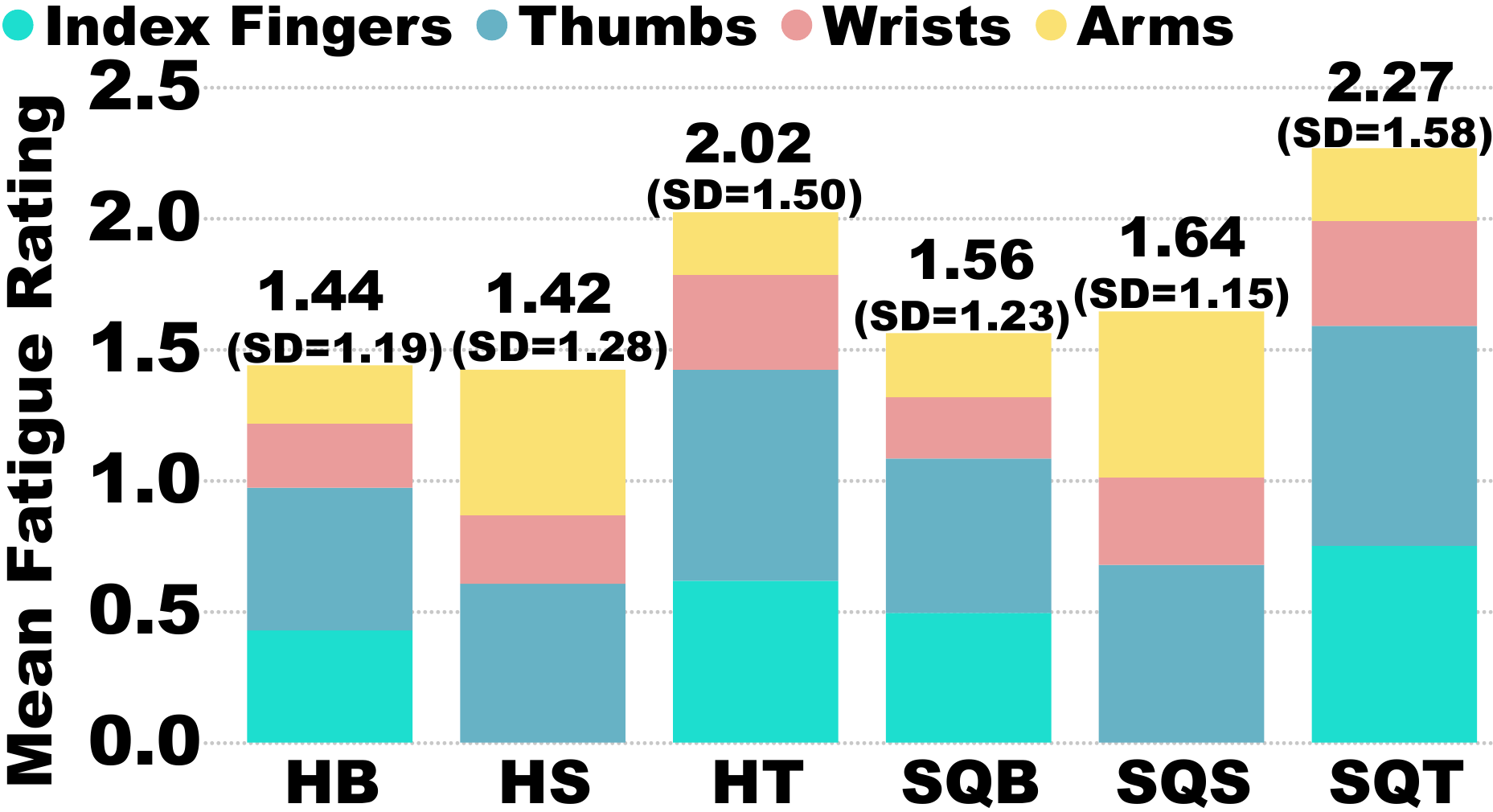}}\\
  \subfloat[Index Fingers]{\includegraphics[width=0.4\textwidth]{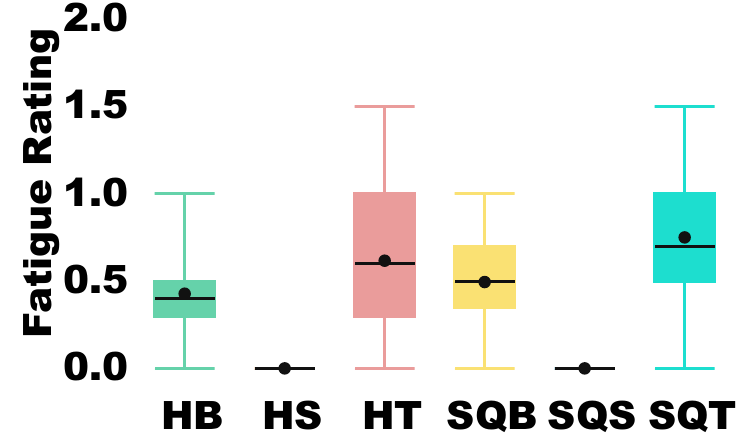}}\
  \hspace{0.9cm}
  \subfloat[Thumbs]{\includegraphics[width=0.4\textwidth]{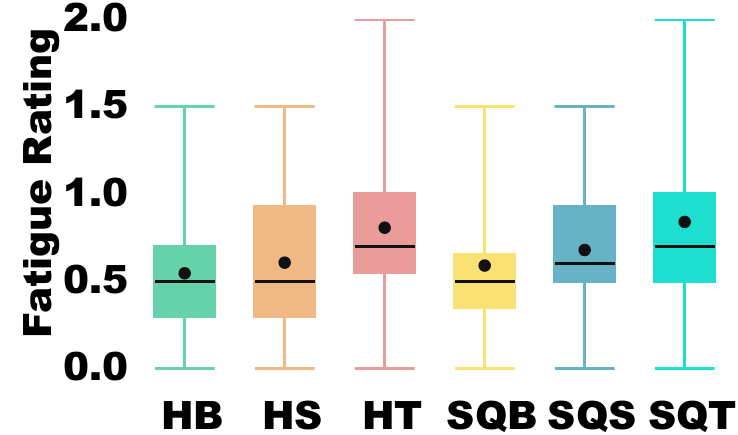}}\\
  \subfloat[Wrists]{\includegraphics[width=0.4\textwidth]{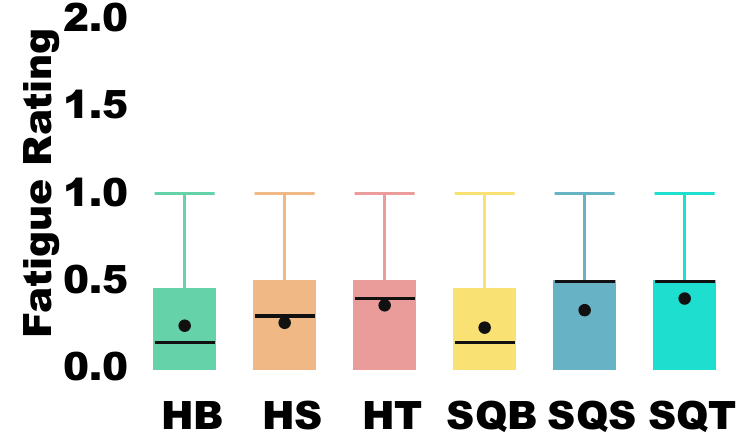}}\
  \hspace{0.9cm}
  \subfloat[Arms]{\includegraphics[width=0.4\textwidth]{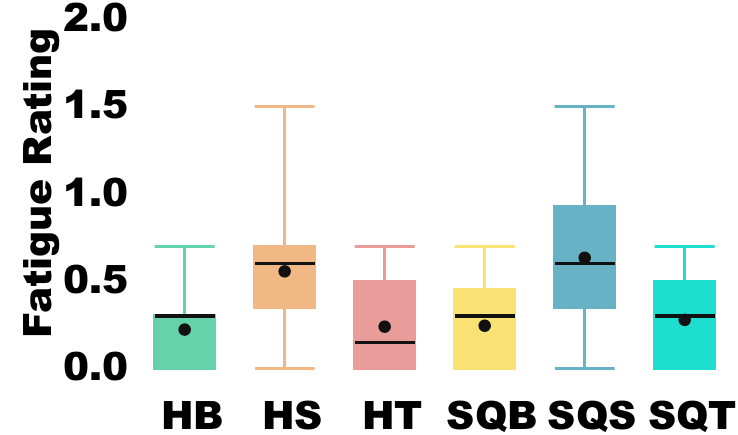}}\\
  \caption{(a) Fatigue ratings in four body parts for different conditions in User Study \Romannum{1}. The mean and standard deviation of accumulated fatigue ratings are shown at the top of each stack.
  (b-e) Details fatigue ratings in the four body parts. For each box, the black dot shows the mean, and the three horizontal lines (from top to bottom) respectively represent the maximum, median, and minimum. The bottom and top edges of each box are the first and third quartiles.}
  \Description{Results of User Study 1.}
  \label{fig:US1_Results}
\end{figure}

\subsection{Interface and Apparatus}
The testing interface and the device control program of our system were implemented using Unity 2021.3.15f1 with C\# and Arduino IDE 2.3.2 with C++, respectively.
We deployed our system based on the HTC Vive Cosmos Elite HMD.
The user study was carried out on a personal computer (with an Intel Core i7 CPU, 80GB RAM, and an NVIDIA GTX 1080Ti graphics card) running Windows 11.

\subsection{Experimental Design and Procedure}
This user study employed a $3\times2$ within-subjects design by considering two independent factors: key confirmation techniques (Button, Shake, and Tap) and keyboard designs (Hive and SplitQWERTY).
This resulted in six different conditions:
(1) Hive+Button (HB);
(2) Hive+Shake (HS);
(3) Hive+Tap (HT);
(4) SplitQWERTY+Button (SQB);
(5) SplitQWERTY+Shake (SQS);
(6) SplitQWERTY+Tap (SQT).

After welcoming the participants, they were introduced on the experiment purpose, protocols, and questionnaires, and also asked to complete a demographic questionnaire and sign a consent form.
Subsequently, the practice session began, and the participants were instructed to familiarize themselves with the VR environment and practice all six conditions.
During the practice session, we asked participants to type five phrases for each condition.
Following a brief intermission of ten minutes, the experiment was divided into six sessions, each of which examined only one of the aforementioned conditions.
To counterbalance the order of conditions, the participants were randomly divided into three groups, each of which contained six participants.
In each group, the same $6\times6$ balanced Latin square design was employed to determine the order of conditions.

In each experiment session, the participants were instructed to transcribe exactly the same 30 phrases (selected from the MacKenzie phrase set~\cite{MacKenzie}) and were encouraged to make corrections using the backspace.
Following each session, we asked the participants to take off the HMD and fill out questionnaires, including Borg CR10~\cite{Borg}, NASA-TLX~\cite{NASATLX}, and UEQ~\cite{UEQ}.
They were then allowed to take a ten-minute break before the next session began.
At the end of the sixth session, the participants were required to fill out a user preference questionnaire, be interviewed, and provide feedback on their overall experience with different conditions.

\subsection{Evaluation Metrics}
We employed the Borg CR10 scale~\cite{Borg} to evaluate the low-level physical fatigue of the participants and NASA-TLX~\cite{NASATLX} to measure different dimensions of workload (mental demand, physical demand, temporal demand, own performance, effort, and frustration).
Typing efficiency was assessed in terms of WPM~\cite{WPMTER} and the total error rate (TER)~\cite{TER} that includes both the corrected error rate (CER, errors made but corrected during the text entry task) and the not corrected error rate (NCER, errors made but not corrected in the final transcribed text).
Moreover, trackball movements were captured in the virtual environment and converted into physical units (mm) for estimating the real-world thumb moving distances.
We also collected user experience ratings by using the user experience questionnaire (UEQ)~\cite{UEQ} to capture subjective impressions of each condition from participants.
Finally, participants rated each condition independently on a 5-point Likert scale (1: least preferred; 5: most preferred) to indicate their overall preference.

\subsection{Results}
We first detected outliers in the collected data based on the interquartile range ($\text{IQR}=Q_3−Q_1$, where $Q_1$ and $Q_3$ are the first and third quartiles, respectively).
Any values smaller than the lower bound ($Q_1-3\times\text{IQR}$) or larger than the upper bound ($Q_3+3\times\text{IQR}$) were considered as potential outliers and then corrected using Winsorization.
Extreme values smaller than the lower bound (or larger than the upper bound) were replaced with the 10th (or 90th) percentile.
Note that since all error rates of a participant, including the TER, CER, and NCER, were highly related to each other and computed from the same collected data, we regarded them as a group rather than correcting them separately.
Specifically, we detected outliers respectively for TERs, CERs, and NCERs.
However, when an extreme value smaller than the lower bound (or larger than the upper bound) was found, all error rates of the same participant were replaced with those of the participant whose TER is the 10th (or 90th) percentile.

We then employed a two-way repeated measures analysis of variance (ANOVA) with Mauchly’s test and the Greenhouse-Geisser adjustment.
Post-hoc pairwise comparisons were performed using t-tests with the Bonferroni correction.
To control the familywise error rate, a threshold of $0.0022$ was applied for determining statistical significance. Furthermore, ANOVA results of significant differences were reported using the exact $p$-values and post-hoc pairwise comparisons were reported only if significance values were smaller than the threshold $0.0022$.

\begin{figure}[t]
  \centering
  \subfloat[WPM]{\includegraphics[width=0.4\textwidth]{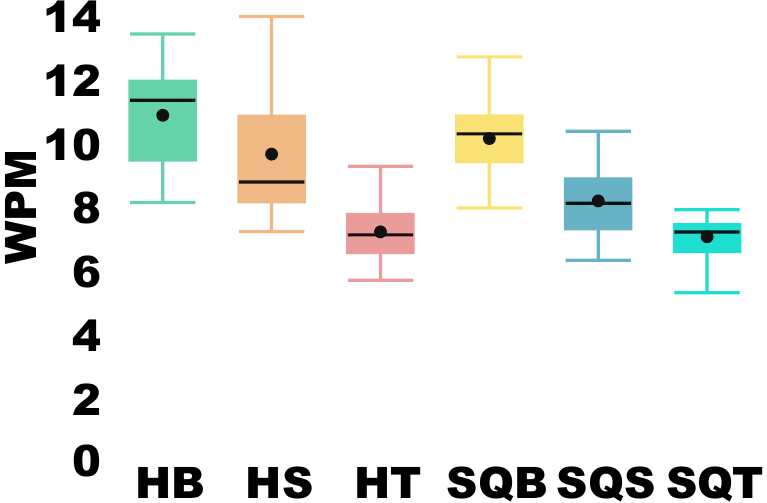}}\
  \hspace{0.9cm}
  \subfloat[TER]{\includegraphics[width=0.4\textwidth]{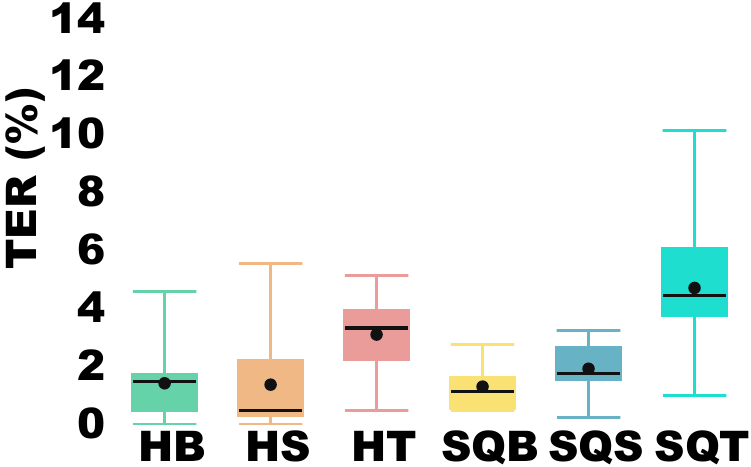}}\\
  \subfloat[CER]{\includegraphics[width=0.4\textwidth]{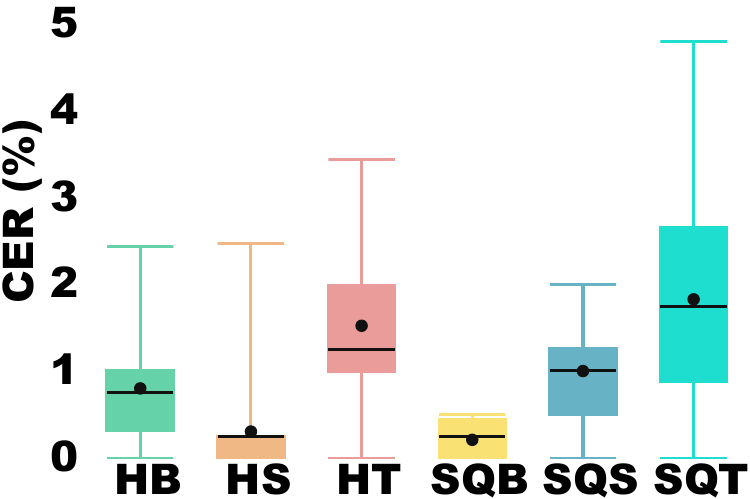}}\
  \hspace{0.9cm}
  \subfloat[NCER]{\includegraphics[width=0.4\textwidth]{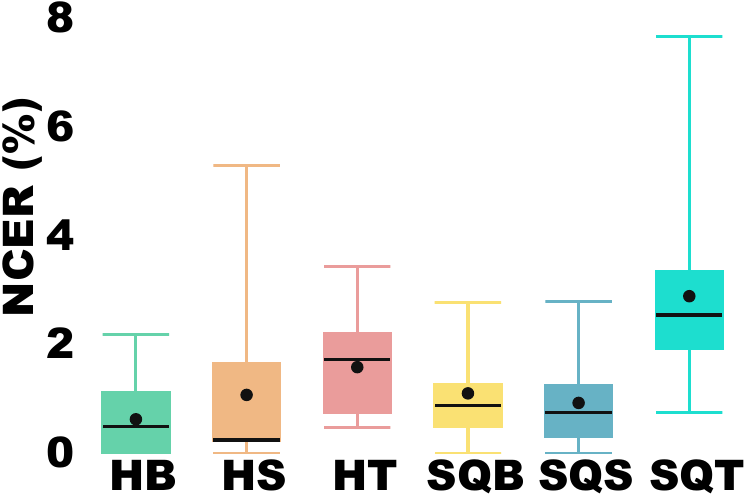}}
  \caption{Box charts of the WPM~\cite{WPMTER}, TERs, CERs, and NCERs~\cite{TER} for different conditions in User Study \Romannum{1}. Refer to Figure~\ref{fig:US1_Results} for the meanings of dots, box edges, and horizontal lines.}
  \Description{Results of User Study 1.}
  \label{fig:US1_Results2}
\end{figure}

\subsubsection{Fatigue}
Figure~\ref{fig:US1_Results}(a) illustrates a stacked chart of the mean fatigue ratings.
Figure~\ref{fig:US1_Results}(b-e) provides more detail on the fatigue ratings in the four body parts.
ANOVA indicated significant effects of confirmation techniques on the fatigue ratings in
index fingers ($F_{2,34}=30.31$, $p=2.83\times10^{-7}$) and
thumbs ($F_{2,34}=12.87$, $p=0.0002$),
but no significant effects for wrists and arms.
No significant effects of keyboard designs were found for index fingers, thumbs, wrists, and arms.
There were also no significant interactions between factors.
Post-hoc pairwise comparisons showed significant differences in the fatigue ratings of index fingers between
HB-HS,
HB-SQS,
HS-HT,
HS-SQB,
HS-SQT,
SQS-HT,
SQS-SQB,
and
SQS-SQT.
However, there were no significant effects on thumbs between all pairs of six conditions.
In general, shaking the device to confirm the selection was significantly different from the others in terms of the fatigue ratings of index fingers.

\subsubsection{Typing Efficiency}
Figure~\ref{fig:US1_Results2}(a) and Figure~\ref{fig:US1_Results2}(b) illustrate the WPM and TERs for the six tested conditions.
For typing speed, ANOVA yielded significant effects of
confirmation techniques ($F_{2,34}=49.52$, $p=2.45\times10^{-9}$)
and
keyboard designs ($F_{1,17}=16.17$, $p=8.87\times10^{-4}$),
but no significant interactions between factors. 
Furthermore, post-hoc pairwise comparisons revealed significant differences in WPM between
HB-HT, HB-SQS, HB-SQT,
HS-HT, HS-SQT,
HT-SQB,
SQB-SQS, and SQB-SQT.

As for TERs, ANOVA showed significant effects of confirmation techniques ($F_{2,34}=30.31$, $p=1.94\times10^{-7}$), but no significant differences for keyboard designs and interactions between factors.
Furthermore, post-hoc pairwise comparisons showed significant differences between
HB-HT, HB-SQT,
HS-SQT,
HT-SQB,
and
SQB-SQT.

Figure~\ref{fig:US1_Results2}(c) and Figure~\ref{fig:US1_Results2}(d) demonstrate the box charts of CERs and NCERs for different conditions.
For CERs, ANOVA indicated significant effects of
confirmation techniques ($F_{2,34}=21.12$, $p=3.05\times10^{-5}$)
and
interactions between factors ($F_{2,34}=8.84$, $p=1.08\times10^{-3}$),
but no significance for keyboard designs.
Furthermore, post-hoc pairwise comparisons showed significant differences in the CER between
HS-HT,
HT-SQB,
and
SQB-SQT.

As for NCERs, ANOVA yielded significant differences in confirmation techniques ($F_{2,34}=13.40$, $p=0.0001$), but no significance for keyboard designs and interactions between factors.
Moreover, post-hoc pairwise comparisons only revealed significant differences between HB-SQT.

\begin{figure}[t]
  \centering
  \subfloat[TMD]{\includegraphics[width=0.4\textwidth]{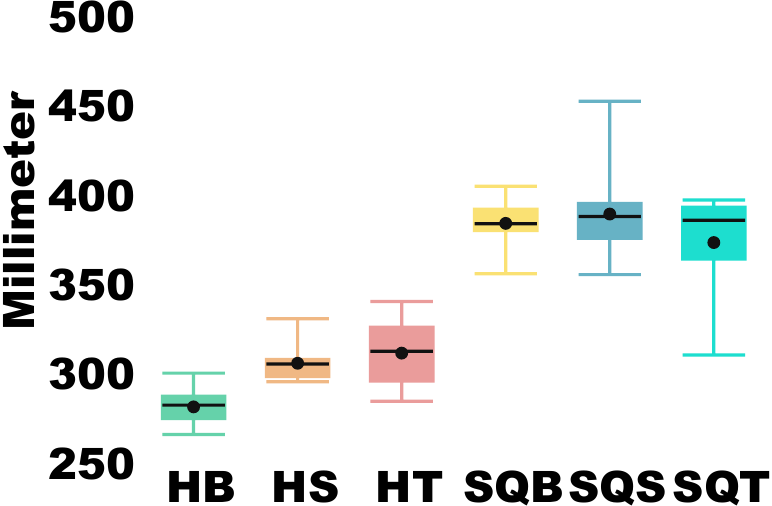}}\
  \hspace{0.9cm}
  \subfloat[Preference]{\includegraphics[width=0.4\textwidth]{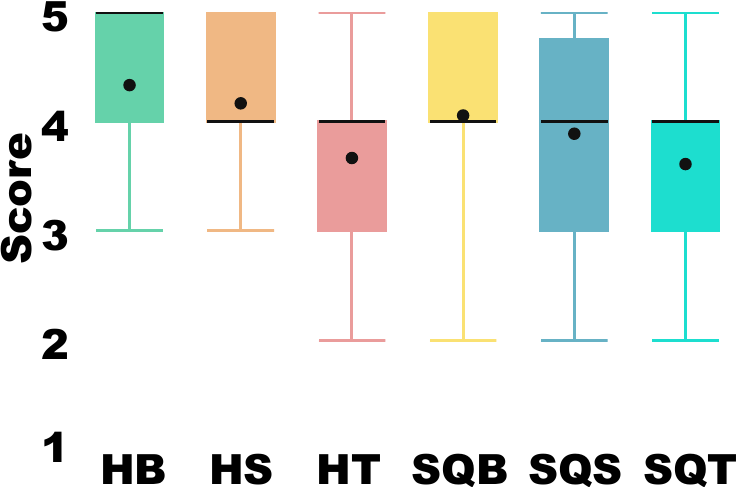}}
  \caption{Box charts of the thumb moving distance (TMD) and user preference (higher is better) for different conditions in User Study \Romannum{1}. Refer to Figure~\ref{fig:US1_Results} for the meanings of dots, box edges, and horizontal lines.}
  \Description{Results of User Study 1.}
  \label{fig:US1_Results3}
\end{figure}

\begin{figure}[ht]
  \centering
  \includegraphics[width=0.99\columnwidth]{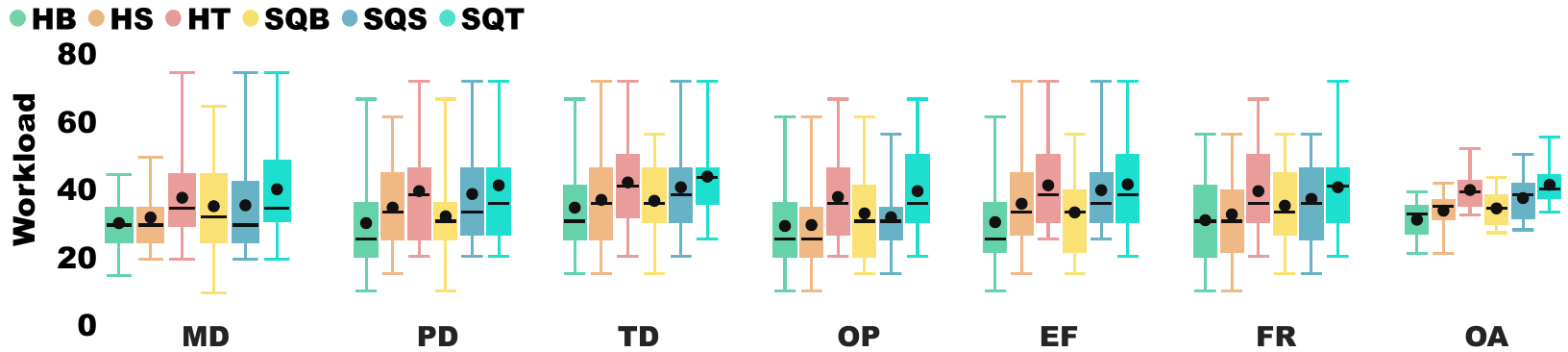}
  \caption{Box chart of NASA-TLX scores, where lower ones are better. The dimensions are mental demand (MD), physical demand (PD), temporal demand (TD), own performance (OP), effort (EF), frustration level (FR), and overall (OA). Refer to Figure~\ref{fig:US1_Results} for the meanings of dots, box edges, and horizontal lines.}
  \Description{Workload scores for User Study 1.}
  \label{fig:US1_NASATlx}
\end{figure}

\subsubsection{Thumb Moving Distance}
Figure~\ref{fig:US1_Results3}(a) shows the box chart of the overall thumb moving distances for the six conditions.
The mean thumb moving distances were
282.65 millimeters ($SD=9.10$) for HB,
307.08 millimeters ($SD=9.24$) for HS,
312.76 millimeters ($SD=17.54$) for HT,
385.43 millimeters ($SD=14.84$) for SQB,
390.66 millimeters ($SD=25.00$) for SQS,
and
374.69 millimeters ($SD=28.84$) for SQT.

ANOVA indicated significant effects of keyboard designs ($F_{1,17}=445.77$, $p=1.23\times10^{-13}$) and interactions between factors ($F_{2,34}=12.06$, $p=0.0003$) on thumb moving distances, but no significant differences for confirmation techniques.
Moreover, post-hoc pairwise comparisons indicated significant differences between all condition pairs except
HS-HT, SQB-SQS, SQB-SQT, and SQS-SQT.

\subsubsection{User Preference}
Figure~\ref{fig:US1_Results3}(b) demonstrates the box chart of user preference scores for the six tested conditions.
The Friedman rank sum test showed no significant effects on user preference ($\chi^2=7.72$, $p=0.17$).
Post-hoc pairwise comparisons using the Wilcoxon signed-rank test also indicated no significant differences among different conditions.

\subsubsection{Workload}
Figure~\ref{fig:US1_NASATlx} illustrates the NASA-TLX scores for the six conditions in detail.
For the overall workload, ANOVA yielded significant effects of confirmation techniques ($F_{2,34}=35.18$, $p=6.89\times10^{-8}$) and keyboard designs ($F_{1,17}=26.20$, $p=8.56\times10^{-5}$), but no significant differences for interactions between factors. 
Moreover, post-hoc pairwise comparisons revealed significant differences between
HB-HT, HB-SQS, HB-SQT, HS-HT, HS-SQT, HT-SQB, and SQB-SQT.
As for individual dimensions, ANOVA showed no significant effects of confirmation techniques, keyboard designs, and interactions between factors.
Furthermore, post-hoc pairwise comparisons also indicated no significance for all pairs of conditions.

\begin{figure}[t]
  \centering
  \includegraphics[width=0.6\columnwidth]{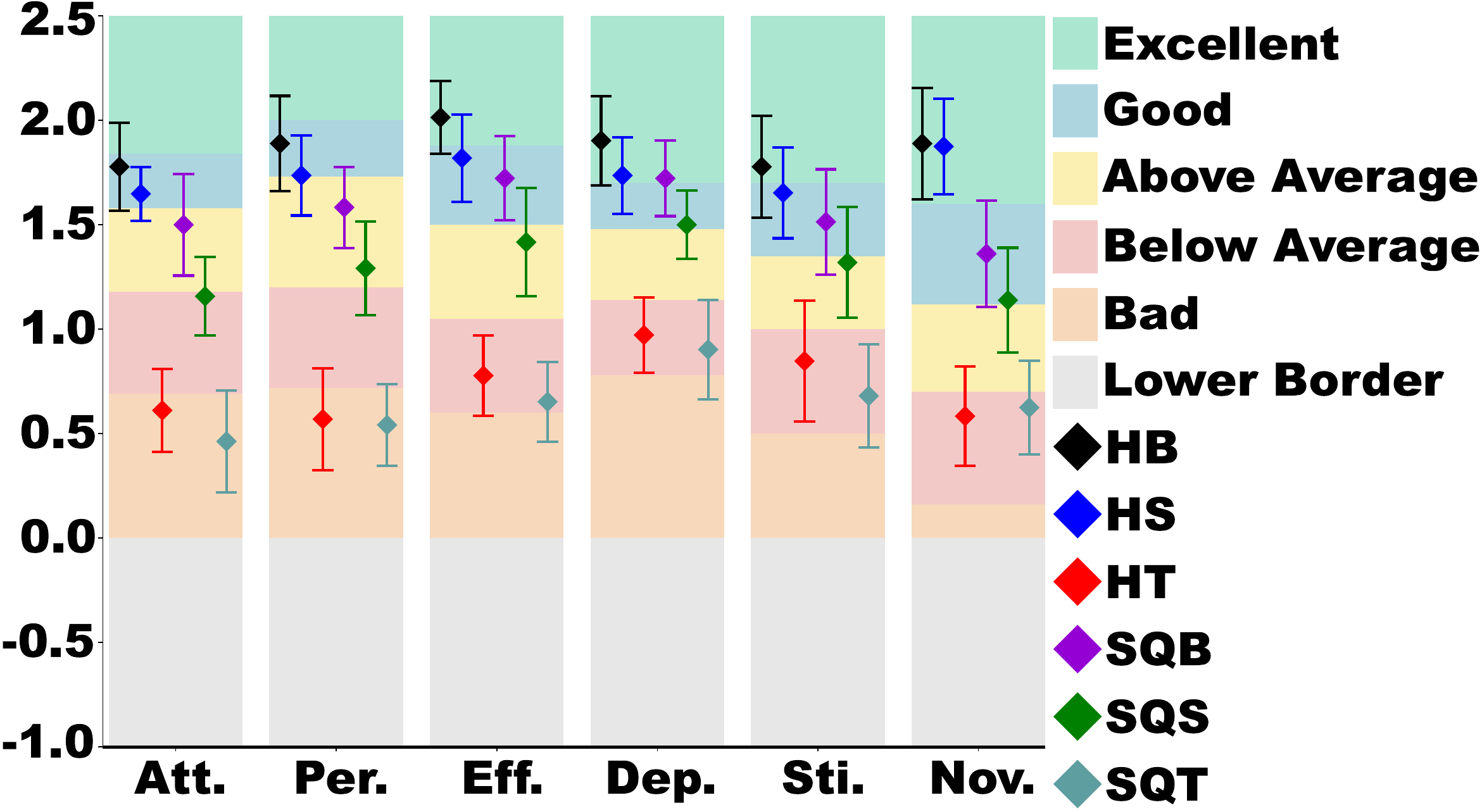}
  \caption{UEQ results in comparison with the established UEQ benchmarks. For each condition, the dot shows the mean, and the two horizontal lines (from top to bottom) respectively indicate the maximum and minimum. The subscales shown are attractiveness (Att.), perspicuity (Per.), efficiency (Eff.), dependability (Dep.), stimulation (Sti.), and novelty (Nov.).}
  \Description{User experience ratings for User Study 1.}
  \label{fig:US1_UEQ}
\end{figure}

\subsubsection{User Experience}
Figure~\ref{fig:US1_UEQ} provides details on the user experience ratings~\cite{UEQ} for the six tested conditions.
ANOVA yielded significant effects of
confirmation techniques ($F_{2,34}=68.66$, $p=3.42\times10^{-10}$)
and keyboard designs ($F_{1,17}=20.55$, $p=2.94\times10^{-4}$)
on attractiveness,
confirmation techniques ($F_{2,34}=70.76$, $p=5.30\times10^{-10}$)
on perspicuity,
confirmation techniques ($F_{2,34}=72.72$, $p=2.15\times10^{-10}$)
on efficiency,
confirmation techniques ($F_{2,34}=66.52$, $p=1.34\times10^{-11}$) and
keyboard designs ($F_{1,17}=13.95$, $p=1.65\times10^{-3}$)
on dependability,
confirmation techniques (\mbox{$F_{2,34}=39.33$}, \mbox{$p=2.67\times10^{-8}$})
on stimulation,
and
confirmation techniques (\mbox{$F_{2,34}=44.98$}, \mbox{$p=2.31\times10^{-8}$}),
keyboard designs (\mbox{$F_{1,17}=18.67$}, \mbox{$p=4.64\times10^{-4}$}),
and
interactions between factors (\mbox{$F_{2,34}=8.60$}, \mbox{$p=1.37\times10^{-3}$})
on novelty.

Moreover, post-hoc pairwise comparisons revealed significant differences in the ratings of
attractiveness between
HB-HT, HB-SQS, HB-SQT, HS-HT, HS-SQS, HS-SQT, HT-SQB, and SQB-SQT,
perspicuity between
HB-HT, HB-SQT, HS-HT, HS-SQT, HT-SQB, SQB-SQT, and SQS-SQT,
efficiency between
HB-HT, HB-SQT, HS-HT, HS-SQT, HT-SQB, HT-SQS, SQB-SQT, and SQS-SQT,
dependability between
HB-HT, HB-SQT, HS-HT, HS-SQT, HT-SQB, HT-SQS, and SQB-SQT,
stimulation between
HB-HT, HB-SQT, HS-HT, HS-SQT, and SQB-SQT,
and
novelty between
HB-HT, HB-SQT, HS-HT, HS-SQS, HS-SQT, and HT-SQB.
In summary, confirmation techniques significantly affected user experience with respect to all subscales, while keyboard designs had an impact on attractiveness, dependability, and novelty.

\subsection{Discussion}
This user study provides strong evidence to answer RQ1 and RQ3 while partially supporting RQ2.
We found that Hive+Button achieved significantly higher text entry rates than Hive+Tap, SplitQWERTY+Tap, and SplitQWERTY+Shake, and outperformed all the other conditions in terms of thumb moving distances.
Furthermore, conditions employing "Button" as key confirmation technique also yielded significantly lower total error rates than those employing "Tap".

For key confirmation techniques, ANOVA results revealed that pressing the button on ErgoGlide to confirm key selections can significantly improve text entry speed.
Most of the participants also mentioned that pressing the button is more intuitive than shaking and tapping and provides a seamless way of text entry in virtual reality, since it did not require them to additionally move any body parts beyond their thumbs to confirm selections.
Moreover, some participants stated that shaking ErgoGlide for a longer period of time was probably tiring.
This can be further validated by the higher mean fatigue ratings of arms in conditions that employed the "Shake" confirmation technique.
One possible reason for this is the gorilla arm fatigue~\cite{Selection,WhyShakeCauseFatigue}.
However, as expected, shaking ErgoGlide did not introduce any fatigue in index fingers based on the fatigue ratings of participants.
Furthermore, conditions that employed "Tap" for confirmation did not perform as expected.
It had the lowest text entry speed, higher error rates, and lower UEQ ratings, which might be owing to the superiority of gesture-based techniques over tapping~\cite{TapandGesture}.
The participants also rated "Tap" as less preferred than other confirmation techniques.

As for keyboard designs, ANOVA indicated that the hive-like virtual keyboard significantly reduced the overall thumb moving distance.
A possible reason is that the SplitQWERTY keyboard is wider than the Hive (although their button sizes are quite similar).
Thirteen participants specifically mentioned that they felt larger sideways thumb movements by using SplitQWERTY.
It can be further validated by our quantitative results, where the three conditions that employed SplitQWERTY (SQB, SQS, and SQT) had significantly larger thumb moving distances.

In summary, this user study evaluated the influence of different key confirmation techniques and keyboard designs on ergonomics, usability/learnability, and typing efficiency in virtual reality.
We explicitly found that confirmation techniques had significant effects on text entry speed, while keyboard designs had greater impacts on thumb movements.
Therefore, while using ErgoGlide, it is suggested to combine the hive-like keyboard with button pressing.

\section{User Study \Romannum{2}: ErgoGlide Versus FanPad, JoyGlide, and PizzaText}
In this user study, we compare ErgoGlide (Hive+Button) with the state-of-the-art touch surface based method FanPad~\cite{FanPad} (with the non-overlapping FanPad layout) and two thumbstick-based methods PizzaText~\cite{PizzaText} (with the 4-key design) and JoyGlide. JoyGlide is a variant of ErgoGlide with thumbsticks. It employs the same virtual keyboard/interface as ErgoGlide, but relies on the thumbsticks and the trigger of a VR controller to respectively move a cursor to the desired key and confirm the selection. FanPad maps the touchpads of both VR controllers to a split QWERTY layout.
For text entry, a user touches and slides their thumb on the touchpad to select a key and confirm the selection by lifting their thumb.
On the other hand, PizzaText integrates a gamepad with a pizza-like virtual keyboard for text entry.
To efficiently type a character, the keyboard is divided into seven slices, each of which contains four characters.
The final selection is then determined by using gamepad thumbsticks.

The major reasons why we chose these approaches are their primary involvement of thumb movements for text entry and the similarity of operating nature to ErgoGlide.
Note that like the original papers, FanPad~\cite{FanPad} and PizzaText~\cite{PizzaText} were evaluated by respectively using touchpads on VR controllers and thumbsticks on a gamepad in our user study.
Further details on the implementation of FanPad and PizzaText can be found in Figure 1 in the supplemental materials of this paper.

\subsection{Research Questions}
Two major research questions were explored in this user study: 
\begin{itemize}
  \item \AddRQ{RQ4} Will ErgoGlide provide higher text entry speed and accuracy than FanPad, JoyGlide, and PizzaText?
  \item \AddRQ{RQ5} Will ErgoGlide improve ergonomics, usability, and learnability over FanPad, JoyGlide, and PizzaText? It is hypothesized that FanPad, JoyGlide, and PizzaText may easily cause physical fatigue in thumbs, since they require relatively larger and stronger thumb movements than ErgoGlide to input a character.
\end{itemize}

\subsection{Participants, Interface, and Apparatus}
This user study included 20 unpaid participants (8 females and 12 males), between the ages of 20-43 ($M=26.55$, $SD=5.21$).
All were students from different universities with diverse ethnic backgrounds and academic disciplines.
Eighteen participants had VR experience by the time of this user study.
Seven had some experience with using touchpads and gamepads, and five had little experience with trackballs.
Note that eighteen participants had participated in User Study \Romannum{1}.
Furthermore, we utilized the same interface and apparatus as in User Study \Romannum{1}, except that we deployed the HTC Vive Cosmos HMD for ErgoGlide and JoyGlide, and the HTC Vive Cosmos Elite HMD for FanPad (due to the touchpads on the controllers) and PizzaText.

\begin{figure}[t]
  \centering
  \subfloat[Mean Fatigue]{\includegraphics[width=0.45\textwidth]{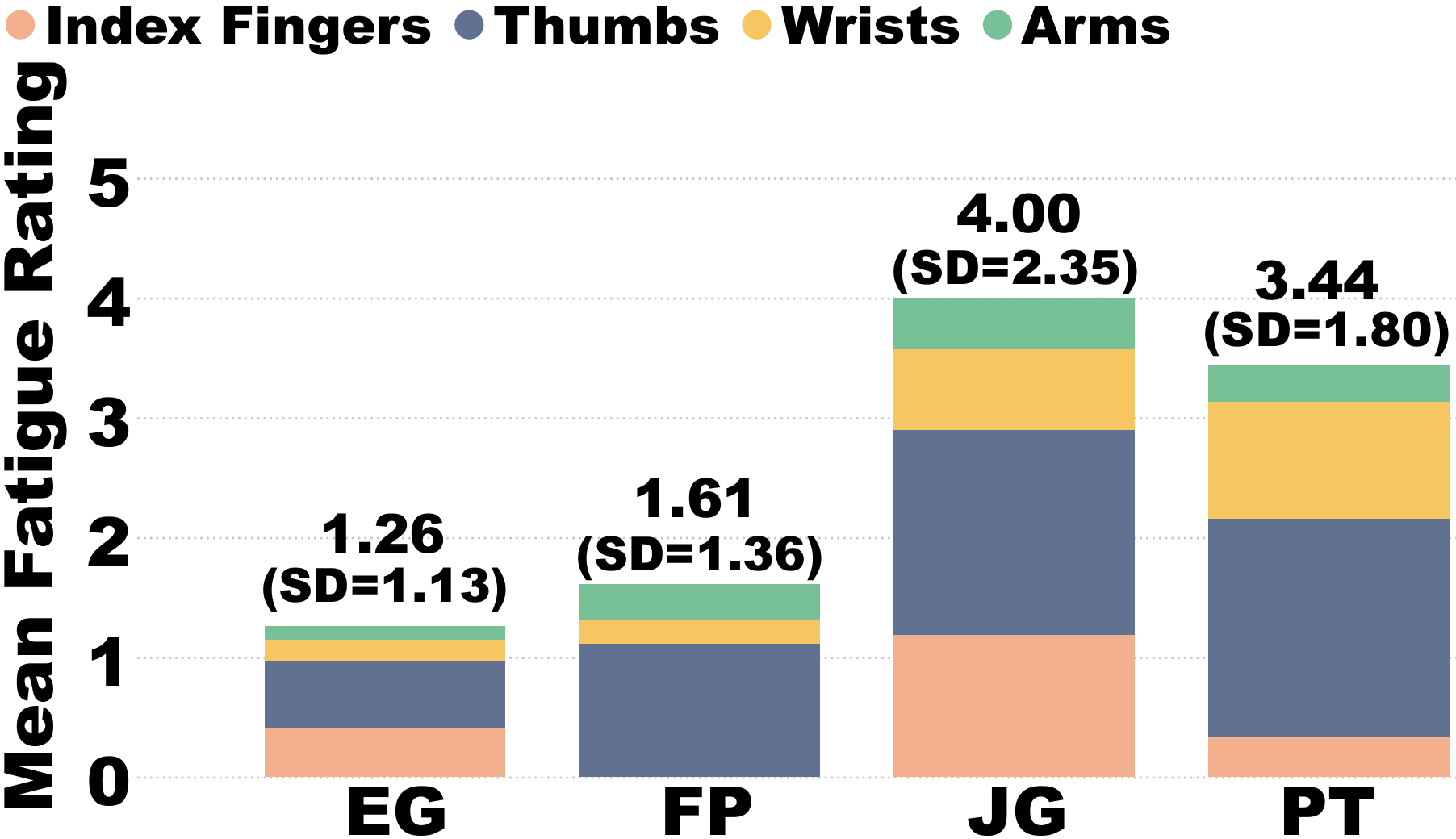}}\\
  \subfloat[Index Fingers]{\includegraphics[width=0.24\textwidth]{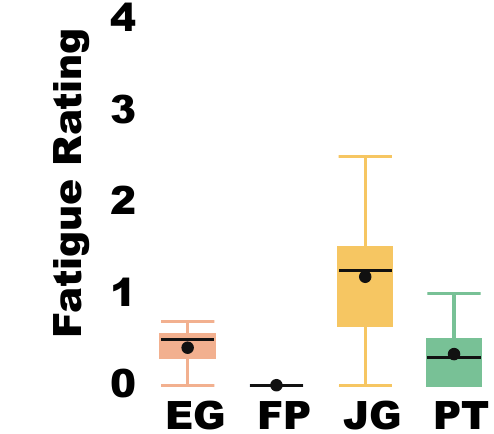}}\
  \subfloat[Thumbs]{\includegraphics[width=0.24\textwidth]{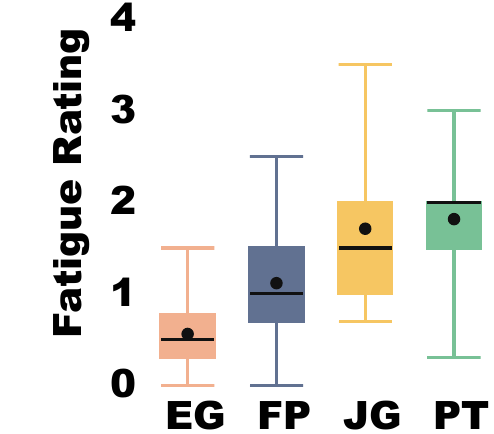}}
  \subfloat[Wrists]{\includegraphics[width=0.24\textwidth]{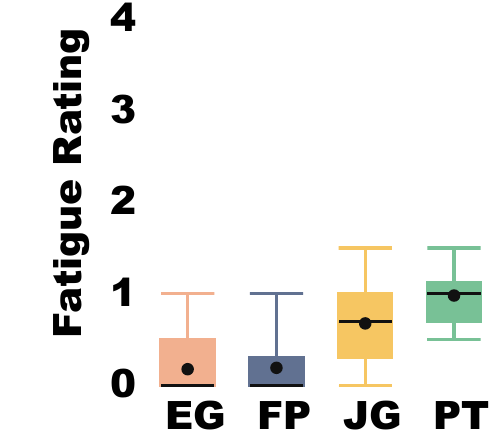}}\
  \subfloat[Arms]{\includegraphics[width=0.24\textwidth]{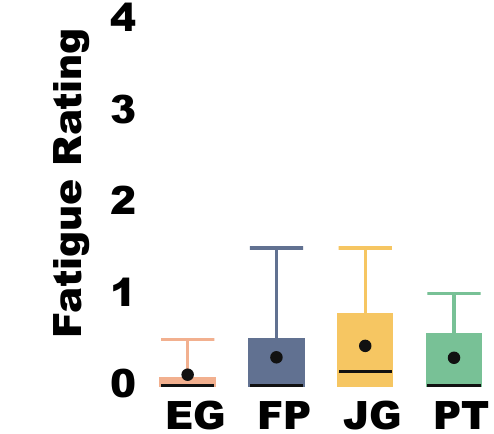}}
  \caption{(a) Fatigue ratings in four body parts for ErgoGlide, FanPad, JoyGlide, and PizzaText. The mean and standard deviation of the accumulated fatigue ratings are shown at the top of each stack. (b-e) Details fatigue ratings in the four body parts. Refer to Figure~\ref{fig:US1_Results} for the meanings of dots, box edges, and horizontal lines. }
  \Description{User Study 2 results.}
  \label{fig:US2_Results}
\end{figure}

\subsection{Experimental Design and Procedure}
Before the user study, the participants were instructed about the objectives and the whole procedure.
They were required to fill out a demographic questionnaire and sign a consent form.
Subsequently, the participants were familiarized with the VR environment and instructed on how to use different text entry methods.
They were then allowed to practice ErgoGlide, FanPad, and JoyGlide for ten minutes each, and PizzaText for fifteen minutes.
After that, during a 10 minute intermission, the participants were briefed on the fatigue (Borg CR10), workload (NASA-TLX), user experience (UEQ), and preference questionnaires.

For each participant, the user study was conducted in four subsequent experimental sessions, each of which evaluated one text entry method.
Similar to User Study \Romannum{1}, the participants were randomly split into five equal groups.
The order of methods/sessions was counterbalanced by applying the $4\times4$ balanced Latin square design to each group.

In each session, all participants were instructed to type exactly the same 30 phrases (selected from the MacKenzie phrase set~\cite{MacKenzie}) as quickly and accurately as possible to minimize confounds.
They were also encouraged to make corrections using the backspace.
Each session lasted approximately 10-15 minutes, depending on the typing speed of the participants.
After each session, the participants were asked to remove the HMD, fill out questionnaires, including Borg CR10, NASA-TLX, and UEQ, and were given an intermission of up to ten minutes between sessions.
To limit response bias, the process of filling out questionnaires was not under observation.
At the end of the fourth session, the participants were required to rate the four methods (on a five-point Likert scale), be interviewed about their experience, and encouraged to make additional comments.

\subsection{Results}
The collected data were cleaned using the same statistical method as in User Study \Romannum{1} and analyzed using a one-way repeated measures ANOVA.
Post-hoc comparisons were performed using pairwise t-tests with the Bonferroni correction.
The sphericity violation was detected by Mauchly’s test, and the Greenhouse-Geisser adjustment was employed when necessary.
Moreover, a significance threshold of $0.0023$ was adopted.
Furthermore, ANOVA results of significant effects were reported using the original $p$-values and post-hoc pairwise comparisons were reported only if significance values were lower than the threshold $0.0023$. In subsequent subsections, EG refers to ErgoGlide, FP to FanPad, JG to JoyGlide, and PT to PizzaText.

\begin{figure}[t]
  \centering
  \subfloat[WPM]{\includegraphics[width=0.24\textwidth]{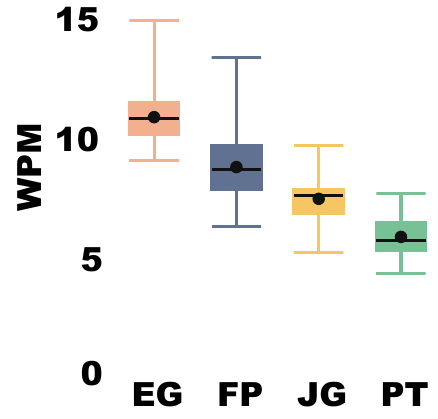}}\
  \subfloat[TER]{\includegraphics[width=0.24\textwidth]{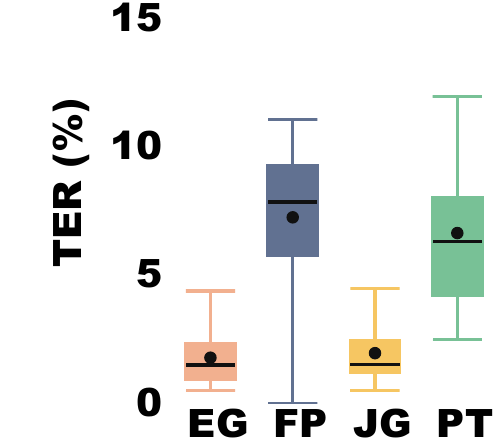}}\
  \subfloat[CER]{\includegraphics[width=0.24\textwidth]{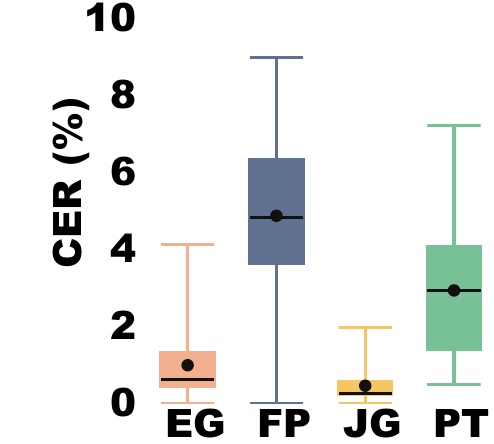}}\
  \subfloat[NCER]{\includegraphics[width=0.24\textwidth]{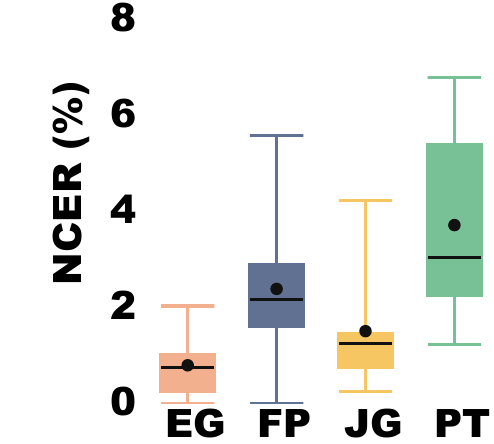}}\\
  \caption{Box charts of the WPM, TERs, CERs, and NCERs for ErgoGlide, FanPad, JoyGlide, and PizzaText. Refer to Figure~\ref{fig:US1_Results} for the meanings of dots, box edges, and horizontal lines. }
  \Description{User Study 2 results.}
  \label{fig:US2_Results2}
\end{figure}

\subsubsection{Fatigue}
Figure~\ref{fig:US2_Results}(a) demonstrates a stacked chart of the mean fatigue ratings and Figure~\ref{fig:US2_Results}(b-e) provides in-depth details on the fatigue ratings in the four body parts.
ANOVA revealed significant differences in the fatigue ratings in
index fingers ($F_{3,57}=34.07$, $p=2.32\times10^{-8}$),
thumbs ($F_{3,57}=43.65$, $p=7.78\times10^{-12}$),
and
wrists ($F_{3,57}=38.50$, $p=8.86\times10^{-9}$),
but no significant differences in arms.

Furthermore, post-hoc pairwise comparisons indicated significant effects on
index fingers between EG-FP, EG-JG, FP-JG, FP-PT, JG-PT,
thumbs between EG-FP, EG-JG, EG-PT, FP-JG, FP-PT,
and
wrists between EG-JG, EG-PT, FP-PT.
Overall, EG exhibited significantly lower index fingers fatigue than JG, thumbs fatigue than all other methods, and wrists fatigue than JG and PT.

\subsubsection{Typing Efficiency}
Figure~\ref{fig:US2_Results2}(a) and Figure~\ref{fig:US2_Results2}(b) reveal the WPM and TER results of different methods.
The mean text entry rates were
10.90 WPM ($SD=1.32$) for EG,
8.82 WPM ($SD=1.65$) for FP,
7.50 WPM ($SD=1.01$) for JG,
and
5.91 WPM ($SD=0.91$) for PT.
The corresponding mean TERs were
1.79\% ($SD=1.09$),
7.27\% ($SD=2.71$),
1.97\% ($SD=1.29$),
and
6.65\% ($SD=2.87$),
respectively.
ANOVA yielded significant differences in both
WPM ($F_{3,57}=69.39$, $p=8.36\times10^{-15}$)
and
TERs ($F_{3,57}=39.79$, $p=6.45\times10^{-11}$).
Post-hoc pairwise comparisons showed significant differences in the WPM between
EG-FP,
EG-JG,
EG-PT,
FP-PT,
JG-PT,
and in the TERs between
EG-FP,
EG-PT,
FP-JG,
JG-PT.
In general, EG is significantly faster than all other methods and more accurate than FP and PT.

Figure~\ref{fig:US2_Results2}(c) and Figure~\ref{fig:US2_Results2}(d) further show the box charts of CERs and NCERs for different methods.
ANOVA yielded significant differences in
CERs ($F_{3,57}=32.92$, $p=8.83\times10^{-9}$)
and
NCERs ($F_{3,57}=19.69$, $p=1.86\times10^{-6}$).
Moreover, post-hoc pairwise comparisons indicated significant differences in
CERs between EG-FP, EG-PT, FP-JG, and JG-PT.
There were also significant effects on NCERs between EG-FP and EG-PT.

\begin{figure}[t]
  \centering
  \includegraphics[width=0.98\columnwidth]{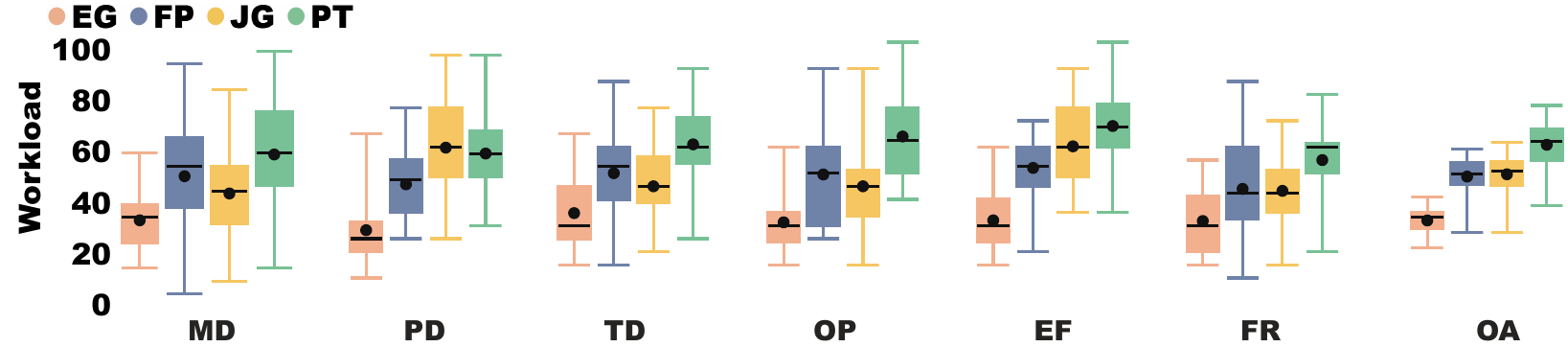}
  \caption{Box chart of NASA-TLX scores, where lower ones are better, for ErgoGlide, FanPad, JoyGlide, and PizzaText. Refer to Figure~\ref{fig:US1_NASATlx} for the meanings of MD, PD, TD, OP, EF, FR, OA and Figure~\ref{fig:US1_Results} for the meanings of dots, box edges, and horizontal lines.}
  \Description{Workload scores for User Study 2.}
  \label{fig:US2_NASATlx}
\end{figure}

\begin{figure}[t]
  \centering
    \subfloat[Preference]{\includegraphics[width=0.33\textwidth]{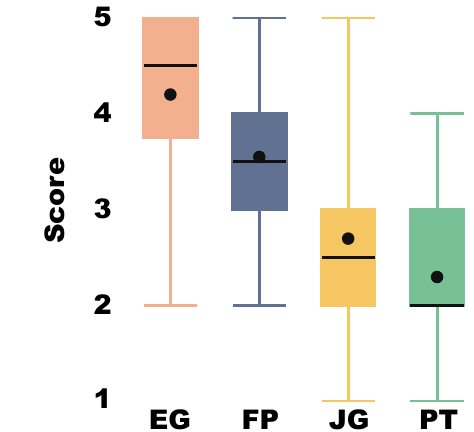}}
    \hspace{0.5cm}
    \subfloat[UEQ]{\includegraphics[width=0.6\textwidth]{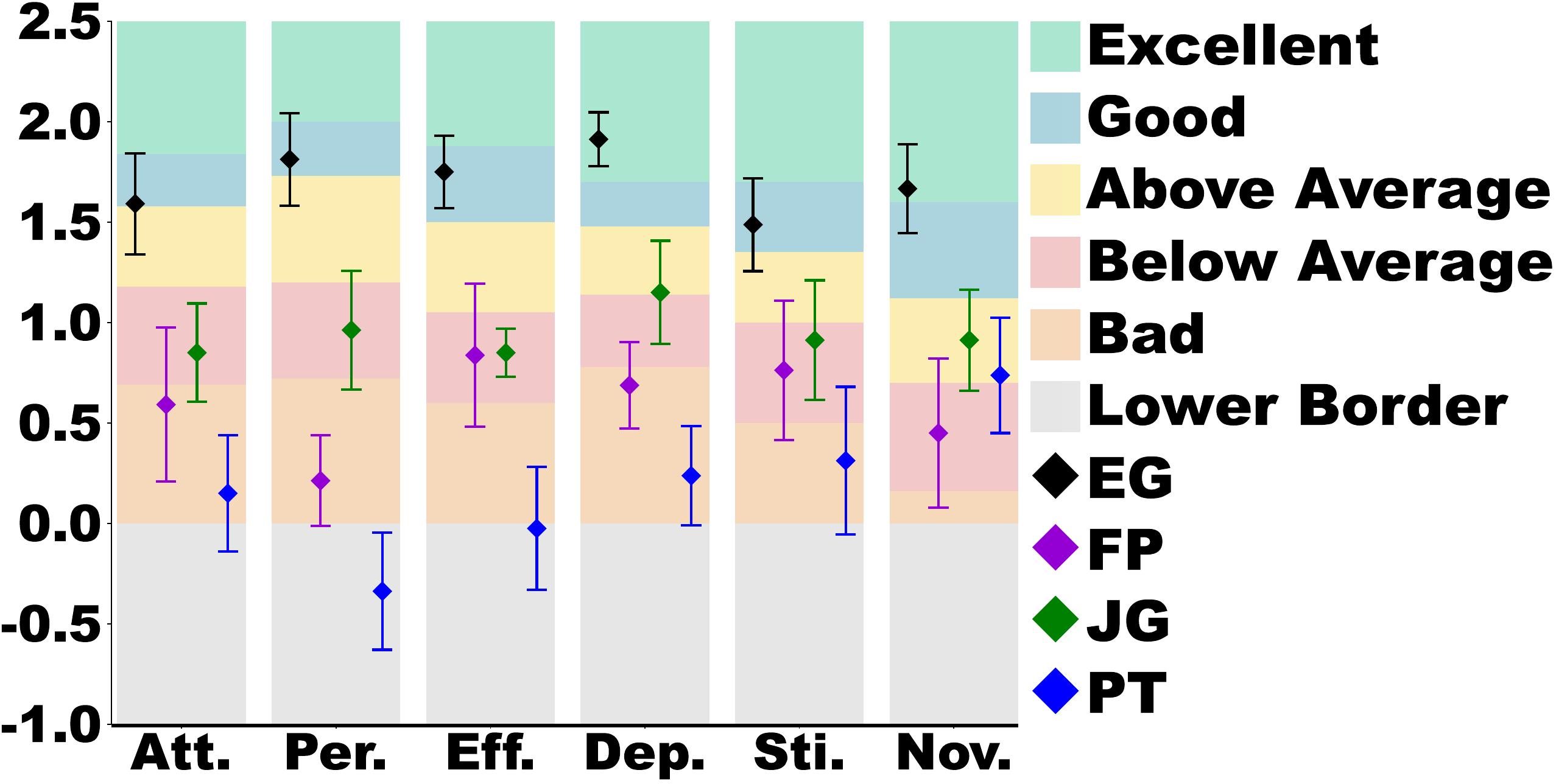}}
  \caption{(a) Box chart of user preference ratings (higher is better) for ErgoGlide, FanPad, JoyGlide, and PizzaText. Refer to Figure~\ref{fig:US1_Results} for the meanings of dots, box edges, and horizontal lines.
  (b) UEQ results in User Study \Romannum{2}. Refer to Figure~\ref{fig:US1_UEQ} for the meanings of dots, horizontal lines, and subscales.}
   \Description{User experience ratings for User Study 2.}
  \label{fig:US2_UEQ_Pref}
\end{figure}

\subsubsection{Workload}
Figure~\ref{fig:US2_NASATlx} illustrates the overall and dimensions NASA-TLX workload scores for different methods.
The mean overall workload scores of EG, FP, JG, and PT were respectively
31.96 ($SD=4.50$),
48.71 ($SD=7.58$),
49.58 ($SD=8.40$),
and
60.83 ($SD=9.46$).
ANOVA indicated significance for
the overall workload ($F_{3,57}=128.90$, $p=8.92\times10^{-16}$)
and all dimensions, including
mental demand ($F_{3,57}=24.19$, $p=3.00\times10^{-7}$),
physical demand ($F_{3,57}=22.80$, $p=1.66\times10^{-7}$),
temporal demand ($F_{3,57}=30.33$, $p=8.22\times10^{-8}$),
performance ($F_{3,57}=36.04$, $p=3.29\times10^{-11}$),
effort ($F_{3,57}=64.29$, $p=2.96\times10^{-14}$),
and
frustration ($F_{3,57}=27.06$, $p=1.56\times10^{-9}$).

Furthermore, post-hoc pairwise comparisons revealed significant differences in the
overall workload between EG-FP, EG-JG, EG-PT, FP-PT, and JG-PT.
For dimensions, significant differences were found 
in mental demand between
EG-FP, EG-PT, and JG-PT,
physical demand between
EG-JG, EG-PT, and FP-PT,
temporal demand between
EG-PT, FP-PT, and JG-PT,
performance between
EG-FP, EG-PT, FP-PT, and JG-PT,
effort between
EG-FP, EG-JG, EG-PT, and FP-PT,
and
frustration between
EG-JG, EG-PT, FP-PT, and JG-PT.
In general, the participants experienced lower workloads while typing with EG, especially in comparison to PT.

\subsubsection{User Preference}
Figure~\ref{fig:US2_UEQ_Pref}(a) illustrates the box chart of user preference ratings for
EG ($M=4.20$, $SD=0.95$),
FP ($M=3.55$, $SD=1.00$),
JG ($M=2.70$, $SD=0.98$),
and
PT ($M=2.30$, $SD=0.86$).
The Friedman rank sum test indicated significant differences in user preference ($\chi^2=39.75$, $p=1.20\times10^{-8}$).
Furthermore, post-hoc pairwise comparisons using Wilcoxon signed-rank tests revealed significance between
EG-JG, EG-PT, and FP-PT.
Overall, the participants preferred EG over thumbstick-based methods (JG and PT) and both EG and FP over PT.

\subsubsection{User Experience}
Figure~\ref{fig:US2_UEQ_Pref}(b) shows the user experience ratings for different methods.
ANOVA yielded significant differences in all subscales of UEQ, including
attractiveness ($F_{3,57}=55.89$, $p=1.10\times10^{-11}$),
perspicuity ($F_{3,57}=128.01$, $p=3.66\times10^{-23}$),
efficiency ($F_{3,57}=62.61$, $p=5.33\times10^{-12}$),
dependability ($F_{3,57}=60.45$, $p=3.20\times10^{-16}$),
stimulation ($F_{3,57}=29.23$, $p=1.02\times10^{-9}$),
and
novelty ($F_{3,57}=33.11$, $p=4.73\times10^{-9}$).

Moreover, post-hoc pairwise comparisons indicated significance for
attractiveness between
EG-FP, EG-JG, EG-PT, and JG-PT,
perspicuity between
EG-FP, EG-JG, EG-PT, FP-JG, FP-PT, and JG-PT, 
efficiency between
EG-FP, EG-JG, EG-PT, FP-PT, and JG-PT,
dependability between
EG-FP, EG-JG, EG-PT, and JG-PT,
stimulation between
EG-FP, EG-JG, EG-PT, and JG-PT,
and novelty between
EG-FP, EG-JG, EG-PT, and FP-JG.
In summary, EG outperformed FP, JG, and PT in all UEQ subscales.

\subsubsection{Discussion}
Our experimental results indicated benefits in terms of ergonomics and efficiency when using ErgoGlide over FanPad, JoyGlide, and PizzaText.
Specifically, ErgoGlide provides significantly higher text entry speed/accuracy (RQ4), lower overall workloads, and better ergonomics, usability, and learnability than the other three methods (RQ5).
One possible reason is that ErgoGlide is both quantitatively and qualitatively more intuitive and user-friendly for the participants, especially novices.
Although the findings reported in the original papers of FanPad~\cite{FanPad} and PizzaText~\cite{PizzaText} differ from those in our user study, this discrepancy is due to differences in user study protocols, such as practice duration, the number of phrases, and total exposure time per method.
Although text entry rates of PizzaText in the original paper were higher than those in our user study, the TERs of PizzaText in our study were lower than those in the original papers.
Other than the different protocols, another possible reason for this is that the participants in our study tended to spend more time accurately correcting typing errors.
We also observed that some participants were quite unfamiliar with FanPad and PizzaText, thus spending more time typing the desired characters.

For ergonomics, the results of the Borg CR10 scale and NASA-TLX respectively showed that ErgoGlide can significantly reduce thumb fatigue and lead to a lower overall workload than FanPad, JoyGlide, and PizzaText (RQ5).
Furthermore, most of the participants preferred ErgoGlide over FanPad and both ErgoGlide and FanPad over thumbstick-based methods (JoyGlide and PizzaText) in terms of physical fatigue.
These findings can be further validated by the significantly lower thumb fatigue ratings and NASA-TLX scores of ErgoGlide than those of FanPad, JoyGlide, and PizzaText.

As for efficiency and learnability, ErgoGlide showed significantly faster text entry speed than FanPad, JoyGlide, and PizzaText and significantly lower error rates than FanPad and PizzaText.
One possible reason for the better efficiency of ErgoGlide is that it uses continuous cursor control for key selections, whereas FanPad and PizzaText employ discrete cursors.
Continuous cursor approaches have been reported to be superior to discrete ones~\cite{Selection}.
Moreover, some participants also mentioned that selecting a character on the touchpad edges is not as easy as selecting one located in the middle of a touchpad.
This may be due to the limited interaction area inherent to touchpad-based methods~\cite{FanPad,PrinType,Velcro}.
Note that 18 out of 20 participants had prior experience with ErgoGlide from User Study \Romannum{1}, which might have introduced an experience bias.
Furthermore, there were similar tendencies in the quantitative data, where NASA-TLX overall and dimension scores, especially effort indicated significantly lower scores for ErgoGlide.

As for usability, ErgoGlide achieved the highest user experience and preference scores, followed by FanPad, JoyGlide, and then PizzaText.
The user study results demonstrated similar tendencies and patterns to those observed for ergonomics and efficiency.
Most participants stated that text entry with ErgoGlide is more intuitive and ergonomic than FanPad, JoyGlide, and PizzaText.
Additionally, thumbstick-based methods, especially PizzaText, may be the worst methods, which is significantly different from ErgoGlide, with higher workload and fatigue ratings in two body parts (thumbs and wrists), slower text entry speed, higher total error rates, and lower usability ratings and preference scores.
This is likely due to the texting thumb problem~\cite{Selection} and the superior performance advantage of the trackballs over thumbsticks~\cite{TrackballSup}.

Note that during the practice session, each participant experienced ErgoGlide, FanPad, and JoyGlide for ten minutes each, and PizzaText for fifteen minutes.
We observed that the participants spent more time getting familiar with PizzaText, but still could not effectively master it during the corresponding experiment session.
However, with more practice time (for example, two hours), novices were able to type as fast as 8.59 WPM based on the user study results of PizzaText~\cite{PizzaText}.
This finding also reveals that the learning curve of ErgoGlide is significantly lower than that of PizzaText.

\begin{figure}[t] 
  \centering
  \subfloat[WPM]{\includegraphics[width=0.35\columnwidth]{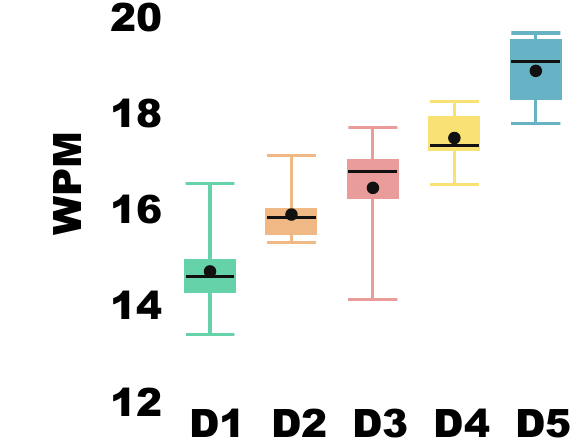}}\
  \hspace{0.9cm}
  \subfloat[TER]{\includegraphics[width=0.35\columnwidth]{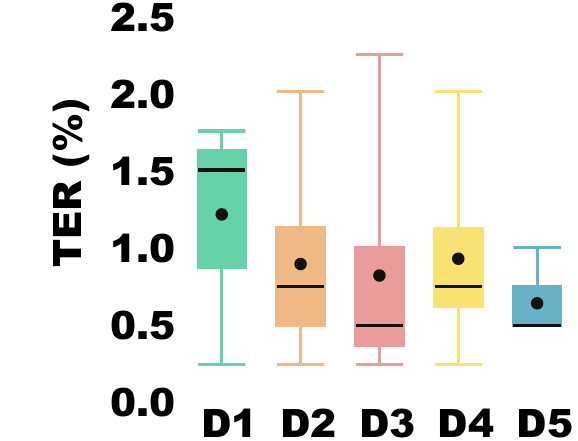}}\\
  \subfloat[CER]{\includegraphics[width=0.35\columnwidth]{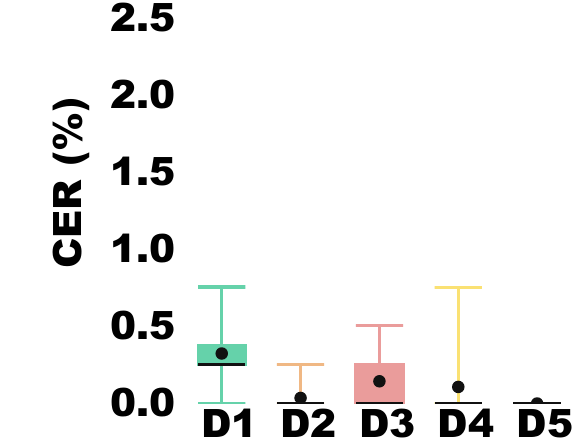}}\
  \hspace{0.9cm}
  \subfloat[NCER]{\includegraphics[width=0.35\columnwidth]{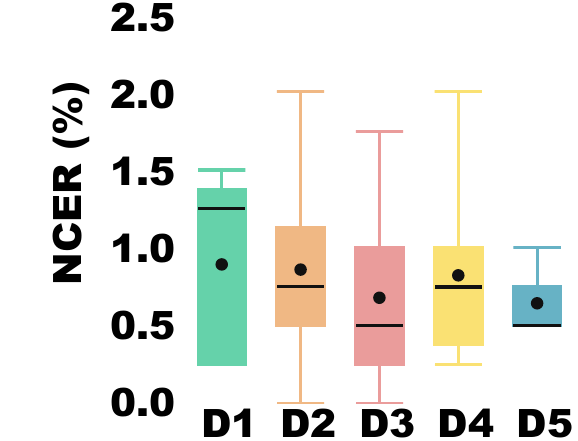}}
  \caption{Box charts of the WPM, TERs, CERs, and NCERs across five days (D1-D5).}
  \Description{Text entry speed and different kind of error rates for ErgoGlide in User Study 3.}
  \label{fig:US3_BoxCharts}
\end{figure}

\section{User Study \Romannum{3}: Performance Evaluation of \mbox{ErgoGlide}}
We conducted this user study to evaluate the impact of training sessions on the performance of ErgoGlide. Specifically, we aimed to investigate how text entry speed and accuracy would be improved over a five-day training period.

\subsection{Research Questions}
This user study addressed the following research question:
\begin{itemize}
  \item \AddRQ{RQ6} Will practice significantly improve text entry speed and accuracy for ErgoGlide?
\end{itemize} 

\subsection{Participants, Interface, and Apparatus}
Seven participants (3 females and 4 males), between the ages of 22-29 ($M=24.86$, $SD=2.67$), from User Study \Romannum{2} voluntarily participated in this user study.
Note that their mean text entry rates were the highest seven in User Study \Romannum{2}.
The same interface and apparatus from User Study \Romannum{1} were employed.

\subsection{Experimental Design and Procedure}
This user study was conducted in five sessions, each of which was run on a separate day.
Prior to the first session, the participants were informed about experimental objectives and procedure, and were required to sign a consent form and fill out a demographic questionnaire.
During each session, the participants were instructed to first practice ErgoGlide for ten minutes, then given a ten-minute break, and finally required to type exactly the same 30 phrases (selected from the MacKenzie phrase set~\cite{MacKenzie}) as quickly and accurately as possible.
Note that the participants were encouraged to make corrections using the backspace during the experiment, but were not asked to fill out any questionnaires in this user study.

\begin{figure}[t] 
  \centering
  \subfloat[WPM]{\includegraphics[width=0.45\columnwidth]{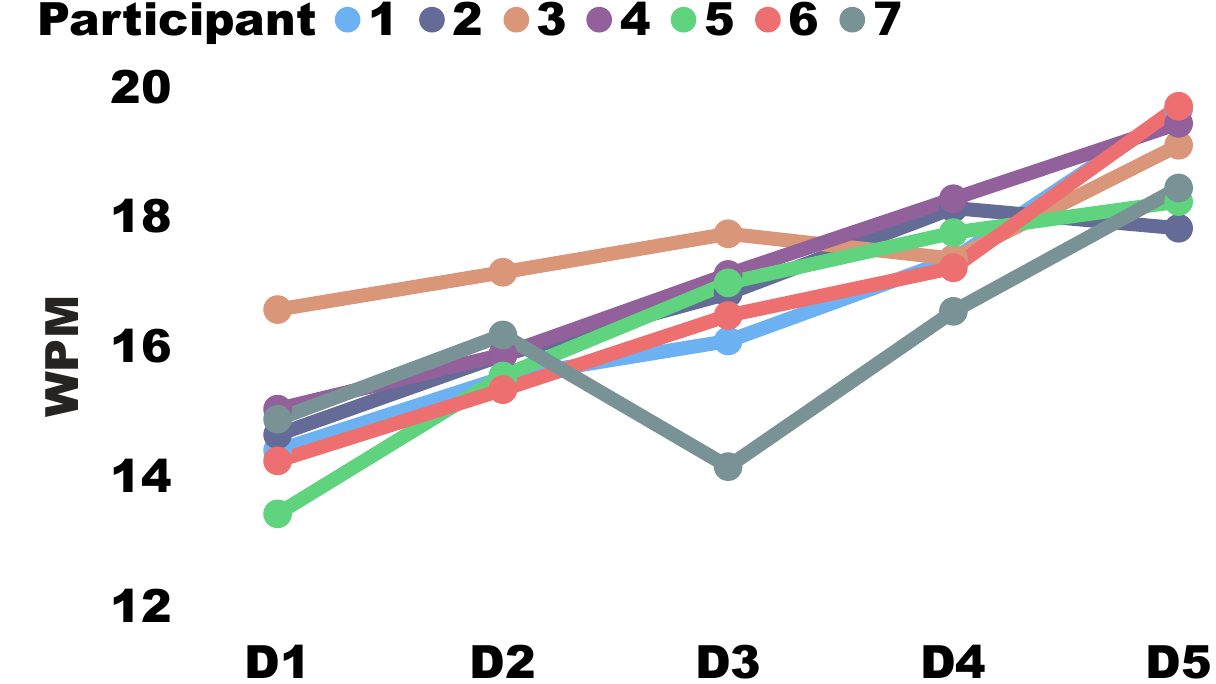}}\
  \subfloat[TER]{\includegraphics[width=0.45\columnwidth]{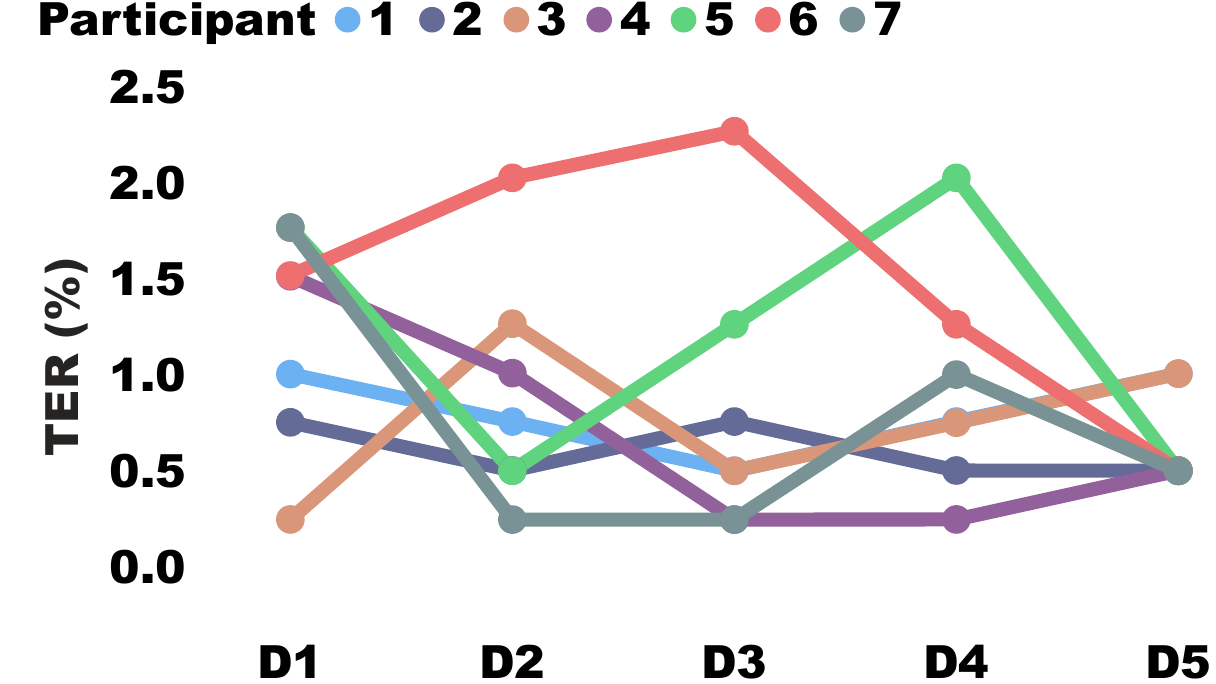}}\\
  \subfloat[CER]{\includegraphics[width=0.45\columnwidth]{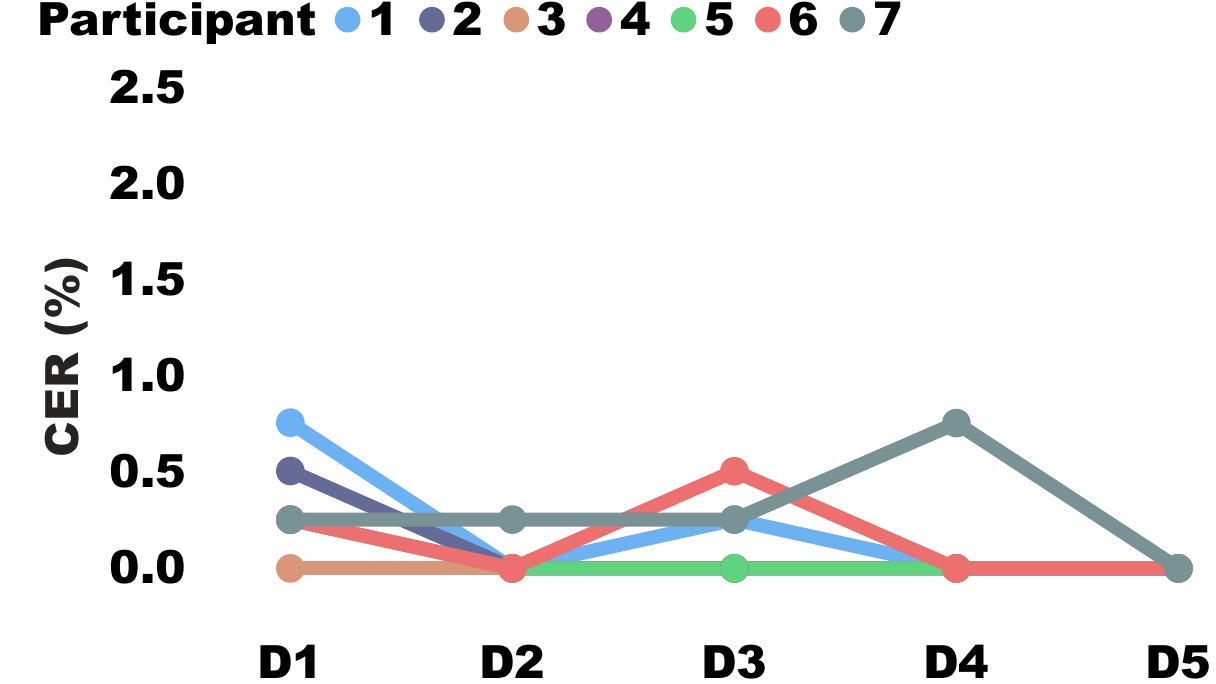}}\
  \subfloat[NCER]{\includegraphics[width=0.45\columnwidth]{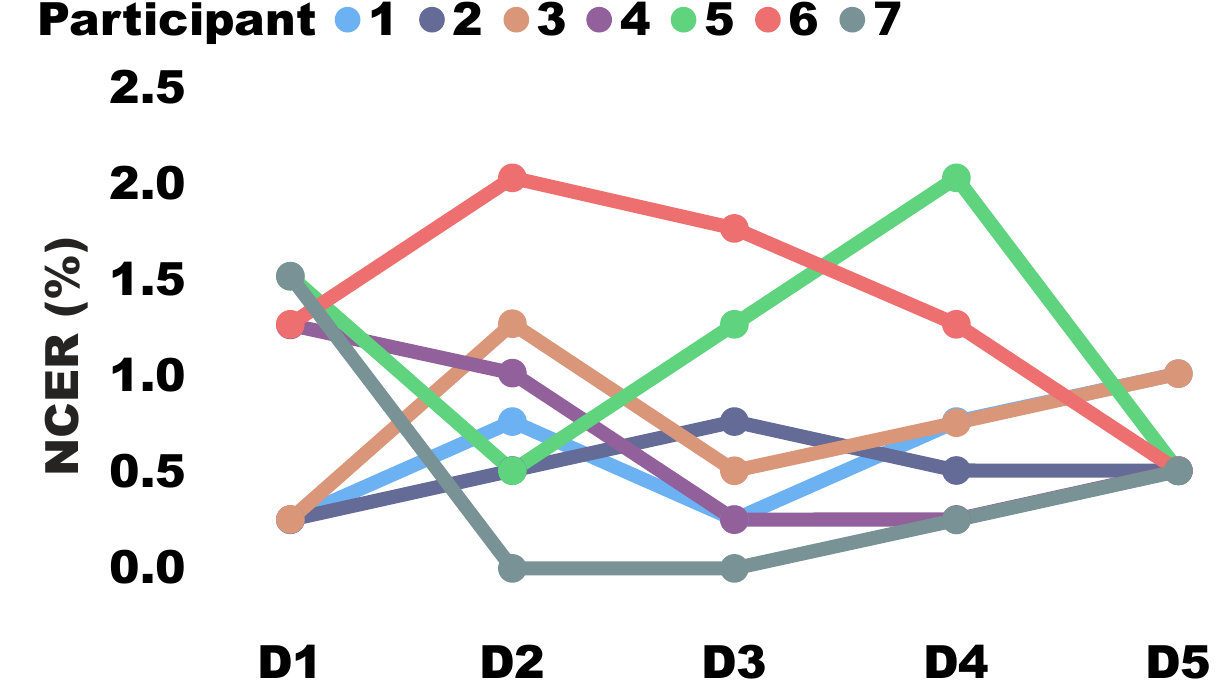}}
  \caption{The WPM, TERs, CERs, and NCERs of individual participants across five days (D1-D5).}
  \Description{The WPM, TERs, CERs, and NCERs of individual participants across five days (D1-D5).}
  \label{fig:US3_LineGraphs}
\end{figure}

\subsection{Results}
The collected data were cleaned and analyzed using the same statistical methods as in User Study \Romannum{2}.
To control the familywise error rate, a significance threshold of 0.01 was applied.

\subsubsection{Typing Efficiency}
Figure~\ref{fig:US3_BoxCharts}(a) and Figure~\ref{fig:US3_LineGraphs}(a) respectively illustrate the box chart of typing speed and individual text entry rates across the five days (D1–D5).
ANOVA revealed a significant effect of training on the WPM ($F_{4, 24}=31.23$, $p=1.44\times10^{-6}$).
Post-hoc pairwise comparisons further indicated significant differences in the WPM between
D1-D2,
D1-D4,
D1-D5,
and
D2-D5.

Figure~\ref{fig:US3_BoxCharts}(b)-(c) and Figure~\ref{fig:US3_LineGraphs}(b)-(c) respectively show the box charts of error rates and the TERs, CERs, and NCERs of individual participants.
ANOVA indicated no significant effects of training on
TERs ($F_{4, 24}=1.15$, $p=0.36$),
CERs ($F_{4, 24}=3.38$, $p=0.03$),
and
NCERs ($F_{4, 24}=0.33$, $p=0.80$).
Moreover, post-hoc pairwise comparisons also revealed no significant differences in TERs, CERs, and NCERs between all pairs of the training days.

\subsubsection{Discussion}
After a cumulative training duration of fifty minutes, the mean text entry rates increased from 14.75 WPM to 18.92 WPM, representing a 28.27\% improvement in text entry speed.
The mean total error rate (TER) decreased from 1.23\% on Day 1 to 0.65\% by Day 5.
Our findings further demonstrate that ErgoGlide achieved higher mean text entry rates than some state-of-the-art touch surface based text entry methods, including HiPad~\cite{HiPad}, discrete and continuous cursor approaches~\cite{Selection}, TipText~\cite{TipText}, and ThumbText~\cite{ThumbText}.
Although the mean text entry rate of ErgoGlide (18.92 WPM) was slightly lower than that of FanPad~\cite{FanPad} (19.73 WPM) by the end of training, the mean total error rate of ErgoGlide (0.65\%) was substantially lower than FanPad (5.56\%).
Furthermore, ErgoGlide also outperformed several other thumbstick-based text entry methods, including PizzaText~\cite{PizzaText} (8.59 WPM), dual thumbsticks with a split-QWERTY layout~\cite{JoyStick4} (7.08 WPM), and EdgeWrite~\cite{JoyStick2} (6.40 WPM), all reported for novice users.
However, with extended training, users have been reported to achieve a text entry rate of 15.85 WPM with PizzaText~\cite{PizzaText} and 10.43 WPM with EdgeWrite~\cite{JoyStick2}.
Although informative, these comparisons should be interpreted with caution, as the testing protocols and experimental environments were not identical across studies.

\section{Conclusion, Limitations, and Future Work}
In this paper, we developed a novel, lightweight, and compact wearable device, namely ErgoGlide, to achieve satisfactory typing accuracy, ergonomics, usability, and learnability for text entry tasks in virtual reality.
ErgoGlide allows a user to rotate the ball for selecting keys on a hive-like virtual keyboard and press the button on the device to enter the selected key, making text entry intuitive and comfortable.
We conducted three user studies to investigate the effectiveness of ErgoGlide and its appropriate configurations. Our first user study revealed that key confirmation techniques had significant impacts on text entry speed, while keyboard design significantly reduced the overall thumb moving distance.
The second user study indicated that the text entry speed/accuracy, ergonomics, and usability/learnability of ErgoGlide are significantly better than those of FanPad (based on touch surfaces), and JoyGlide and PizzaText (based on thumbsticks).
The third user study demonstrated a significant improvement in text entry performance after a short training of fifty minutes.

In summary, ErgoGlide is an ergonomically user-friendly text entry method for applications where users are required to type many words over a medium to long period of time.
To the best of our knowledge, ErgoGlide may be the first wearable trackball-based device that facilitates accurate, easy-to-learn, and ergonomically user-friendly text entry tasks in virtual environments.
ErgoGlide also can be employed across the reality–virtuality continuum to support various XR and Metaverse applications.

While ErgoGlide has been proven to be ergonomic, easy-to-learn, and efficient/accurate, it still has certain limitations.
First, we adopted an optical displacement sensor to measure 2D displacement vectors, such that ErgoGlide is sensitive to wearing orientation.
Second, some participants stated that tapping or shaking may be suitable for different postures. However, during our experiment, we only tested the standing posture. A further in-depth evaluation may be required to validate this finding.
Third, we examined the interoperability on a limited number of systems/platforms. Although the results are consistent across tested systems/platforms, future work will focus on testing ErgoGlide on a larger number of systems/platforms.
To further improve text entry speed, we are interested in exploring more appropriate key confirmation techniques, keyboard designs, and typing methods (e.g. gesture) for ErgoGlide.
Moreover, we also plan to investigate the performance of ErgoGlide in diverse interaction tasks beyond text entry, such as 3D object manipulation, menu navigation, and general cursor control in different XR environments.

\begin{acks}
This work was supported in part by the National Science and Technology Council (formerly the Ministry of Science and Technology) of Taiwan under Grant Numbers
MOST110-2221-E-155-022-MY3,
NSTC112-2221-E-155-022,
and
NSTC113-2221-E-155-045-MY2.
\end{acks}

\bibliographystyle{ACM-Reference-Format}
\bibliography{sample-base}
\end{document}